\documentclass{pasj00}
\usepackage{color}
\color{black}
\tenpoint

\SetRunningHead{K. Sato et al.}{
Suzaku observation of NGC~1550}

\title{
Metallicity of the Fossil Group NGC~1550\\
Observed with Suzku
}
\author{
 Kosuke \textsc{Sato},\altaffilmark{1}
 Madoka \textsc{Kawaharada},\altaffilmark{2}
 Kazuhiro \textsc{Nakazawa},\altaffilmark{3}\\
 Kyoko \textsc{Matsushita},\altaffilmark{4}
 Yoshitaka \textsc{Ishisaki},\altaffilmark{5}
 Noriko \textsc{Y.~Yamasaki},\altaffilmark{2}
and  Takaya \textsc{Ohashi}\altaffilmark{5} 
}
\altaffiltext{1}{
Kavli Institute for Astrophysics and Space Research,
Massachusetts Institute of Technology, \\
77 Massachusetts Ave., Cambridge, MA 02139, USA}
\email{k\_sato@mit.edu}
\altaffiltext{3}{
Institute of Space and Astronautical Science (ISAS),
Japan Aerospace Exploration Agency, \\
3-1-1 Yoshinodai, chuo-ku, Sagamihara, Kanagawa 252-5210}
\altaffiltext{3}{
Department of Physics, University of Tokyo,
 7-3-1 Hongo, Bunkyo, Tokyo 113-0033} 
\altaffiltext{4}{
Department of Physics, Tokyo University of Science,
1-3 Kagurazaka, Shinjuku-ku, Tokyo 162-8601}
\altaffiltext{5}{
Department of Physics, Tokyo Metropolitan University,
 1-1 Minami-Osawa, Hachioji, Tokyo 192-0397}
\Received{2010 August 11}
\Accepted{2010 September 21}
\Published{---}

\begin{document}
\maketitle

\begin{abstract}
We studied the temperature and metal abundance distributions of
the intra-cluster medium (ICM) in a group of galaxies NGC~1550
observed with Suzaku.  The NGC~1550 is classified as a fossil group, 
which have few bright member galaxies except for the central galaxy.  
Thus, such a type of galaxy is important to investigate how the metals
are enriched to the ICM.  With the Suzaku XIS instruments, we directly 
measured not only Si, S, and Fe lines but also O and Mg lines 
and obtained those abundances to an outer region of
$\sim0.5~ r_{180}$ for the first time, and confirmed that the metals 
in the ICM of such a fossil group are indeed extending to a large 
radius. We found steeper gradients for
Mg, Si, S, and Fe abundances, while O showed almost flat abundance
distribution.  Abundance ratios of $\alpha$-elements to Fe were 
similar to those of the other groups and poor clusters.  We
calculated the number ratio of type II to type Ia supernovae for the
ICM enrichment to be $2.9\pm 0.5$ within $0.1~r_{180}$, and the value
was consistent with those for the other groups and poor clusters
observed with Suzaku. We also
calculated metal mass-to-light ratios (MLRs) for Fe, O and Mg
with B-band and K-band luminosities of the member galaxies
of NGC~1550. The derived MLRs were comparable to those of 
NGC~5044 group in the $r<0.1~r_{180}$ region, while those 
of NGC~1550 are slightly higher than those of NGC~5044 
in the outer region.
  
\end{abstract}

\section{Introduction}

\begin{table*}
\caption{Suzaku Observation logs for NGC~1550.}
\label{tab:1}
\begin{tabular}{lccccc} \hline 
Region & Sequence No. & Observation date & \multicolumn{1}{c}{(RA, Dec)$^\ast$} &Exp.&After screening \\
&&&J2000.0 & ksec &(BI/FI) ksec \\
\hline 
center & 803017010 & 2008-08-16T04:27:05 & (\timeform{04h19m47.7s},
 \timeform{+02D24'38''})& 83.3& 82.0/83.3\\
offset & 803018010 & 2008-08-15T02:44:04 & (\timeform{04h20m59.5s},
 \timeform{+02D24'31''})& 41.1& 40.5/40.7\\
\hline\\[-1ex]
\multicolumn{6}{l}{\parbox{0.9\textwidth}{\footnotesize 
\footnotemark[$\ast$]
Average pointing direction of the XIS, written in the 
RA\_NOM and DEC\_NOM keywords of the event FITS files.}}\\
\end{tabular}
\end{table*}

Groups and clusters of galaxies play a key role for investigating 
the formation of the universe and they act as a building blocks 
in the framework of a hierarchical formation of structures. 
The metal abundances of Intra-cluster medium (ICM) in groups and
clusters carry a lot of information in understanding
the chemical history and evolution of groups and clusters.
Recent X-ray observations allow us to measure temperature and metal 
abundance distributions in the ICM based on the spatially resolved spectra.  
A large amount of metals of the ICM are mainly produced by supernovae (SNe) 
in early-type galaxies \citep{arnaud92,renzini93}, 
which are classified roughly as type Ia (SNe~Ia) and type II (SNe~II)\@.
Because Si and Fe are both synthesized in SNe Ia and II, 
we need to know O and Mg abundances,
which are synthesized predominantly in SNe II, in resolving the past
metal enrichment process in ICM by supernovae.  
In order to know how the ICM has been
enriched, we need to measure the amount and distribution of all the
metals from O to Fe in the ICM\@.  

\citet{renzini97} and \citet{makishima01} summarized
iron-mass-to-light ratios (IMLR) with B-band luminosity 
for various objects with ASCA, as a function of
their plasma temperature serving as a measure of the system richness, 
and IMLRs in groups were found to be smaller than those in clusters. 
They also showed that the early-type galaxies released a large
amount of metals which were probably formed through past supernovae 
explosions as shown earlier by \citet{arnaud92}.  
In order to obtain a correct modeling of ICM, 
we need to know the correct temperature and metal abundance profiles
without biases (e.g., \cite{buote00,sanders02}).
Especially for the ICM of cooler systems, such as elliptical galaxies and 
groups of galaxies,
careful analysis are required as mentioned 
in \citet{arimoto97} and \citet{matsushita00}.

The spatial distribution and elemental abundance pattern
of the ICM metals were determined with the large effective area of
XMM-Newton
\citep{matsushita07b,tamura04,boehringer05,osullivan05,sanders06,
deplaa06,deplaa07,werner06,simionescu08}.
On the other hand, the abundance measurements of O and Mg with XMM-Newton 
were possible only for the central regions
of brightest cooling core clusters due to the relatively high 
intrinsic background.
\citet{rasmussen07,rasmussen09} report the Si and Fe profiles of 
15 groups of galaxies observed with Chandra. They suggest that 
the Si to Fe ratios in groups tend to increase with radius, 
and the IMLRs within $r_{500}$ show a positive 
correlation with total group mass (temperature). 
Suzaku XIS can measure all the main elements
from O to Fe, because it realizes lower background level and higher
spectral sensitivity, especially below 1 keV \citep{koyama07}.  Suzaku
observations have shown the abundance profiles of O, Mg, Si, S, and Fe
to the outer regions with good precision for several clusters
\citep{matsushita07a,sato07a,sato08,sato09a,sato09b,tokoi08,komiyama09}.  
Combining the Suzaku results with supernova nucleosynthesis models, 
\citet{sato07b} showed the number ratios of SNe~II to Ia to be ~3.5\@.

NGC~1550 is a S0 galaxy and one of the nearest ($z=0.0124$) X-ray bright
galaxies. The NGC~1550 is also classified as a fossil group 
\citep{jones03}, and an X-ray extended object RX J0419+0225 was first 
discovered by the ROSAT ALL SKY SURVEY from a position centered on 
the NGC~1550 galaxy. From ASCA observation 
\citep{kawaharada03,fukazawa04}, the MLRs with B-band is comparable to 
those of clusters. \citet{sun03} reports the temperature drop at the 
central region and also declines beyond 0.1 times of the virial radius, 
$r_{180}$, with Chandra observation.
\citet{kawaharada09} shows the gas mass and metal mass from 
the temperature and metal abundances observed with XMM-Newton.
In addition, they derived the mass-to-light ratios of O, Si, 
and Fe with near infrared (K-band) luminosity. The resultant 
IMLR within $\sim 200~h_{72}$ kpc exhibits about 2 orders of magnitude 
decrease toward the center. 
NGC~1550, such a fossil group, is a important object 
to investigate how the metals have been enriched to the ICM, 
because of little metal supply from the present-day member galaxies. 

This paper reports on results from Suzaku observations of NGC~1550
out to $30'\simeq 457\; h_{70}^{-1}$~kpc, corresponding to
$\sim 0.47\; r_{180}$.  We use $H_0=70$
km~s$^{-1}$~Mpc$^{-1}$, $\Omega_{\Lambda} = 1-\Omega_M = 0.73$ in this
paper.  At a redshift of $z=0.0124$, $1'$ corresponds to 15.2~kpc,
and the virial radius, $r_{\rm 180} = 1.95\;
h_{100}^{-1}\sqrt{k\langle T\rangle/10~{\rm keV}}$~Mpc
\citep{markevitch98}, is 0.97~Mpc (\timeform{63.5'}) for an average temperature
of $k\langle T\rangle = 1.2$~keV\@.  Throughout this paper we adopt
the Galactic hydrogen column density of $N_{\rm H} = 1.15\times
10^{21}$ cm$^{-2}$ \citep{dickey90} in the direction of NGC~1550\@.
Unless noted otherwise, the solar abundance table is given by
\citet{anders89}, and the errors are in the 90\% confidence region 
for a single interesting parameter.

\begin{figure}
\centerline{
\FigureFile(0.5\textwidth,0.5\textwidth){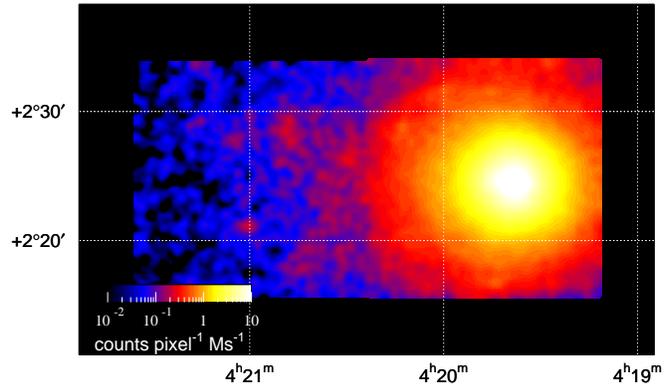}}
\caption{
Combined XIS image in the 0.5--2.0 keV energy range. 
The observed XIS0, 1, and 3 images 
were added on the sky coordinate after removing each calibration source region,
and smoothed with $\sigma=16$ pixel $\simeq 17''$ Gaussian.
Estimated components of extragalactic X-ray background (CXB)
and instrumental background (NXB) were subtracted,
and the exposure was corrected, though vignetting was not corrected.
The white circles show the extracted regions in the spectral fits.
}\label{fig:1}
\end{figure}

\begin{figure*}[hbt]
\begin{minipage}{0.33\textwidth}
\FigureFile(\textwidth,\textwidth){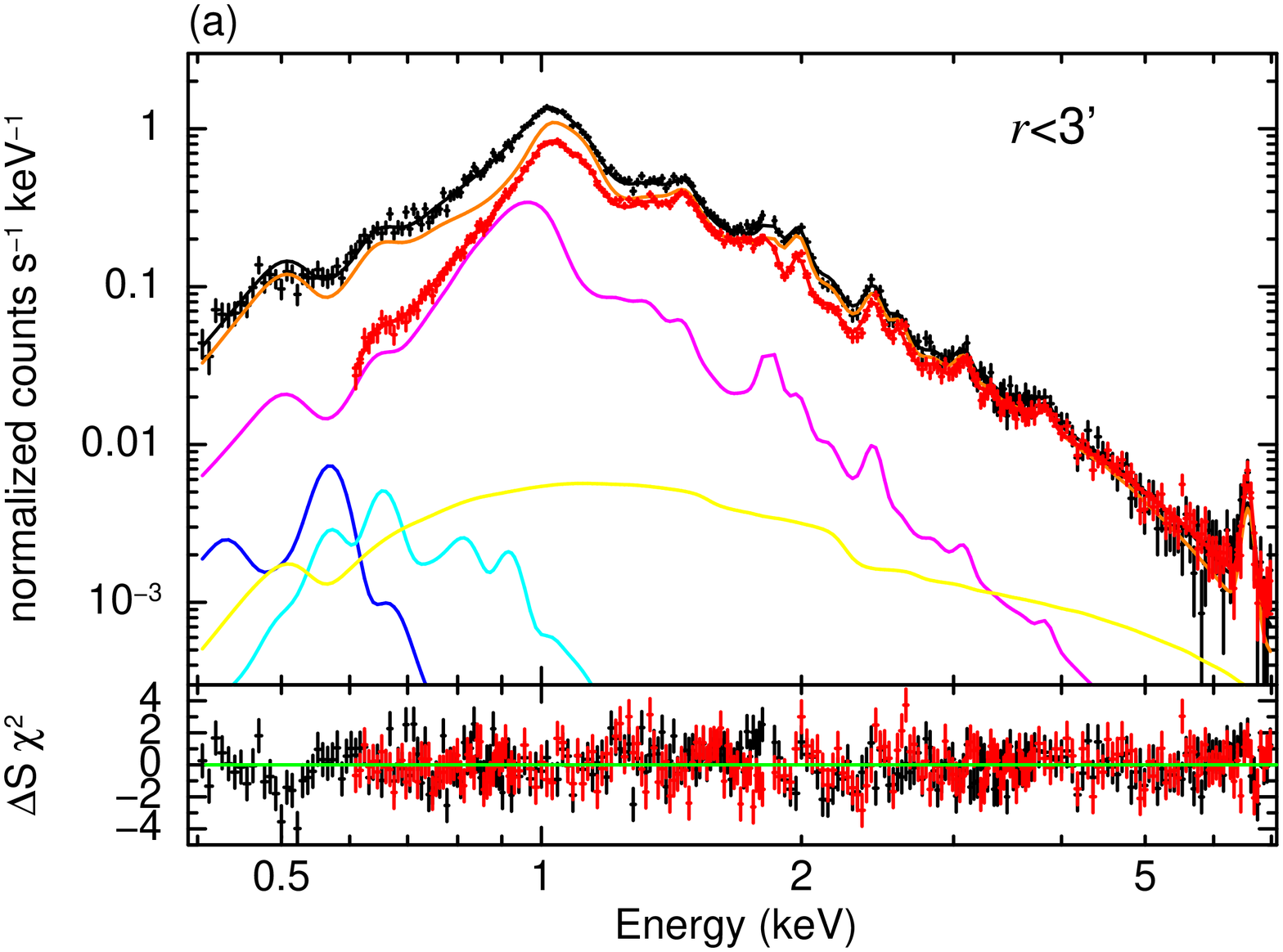}
\end{minipage}\hfill
\begin{minipage}{0.33\textwidth}
\FigureFile(\textwidth,\textwidth){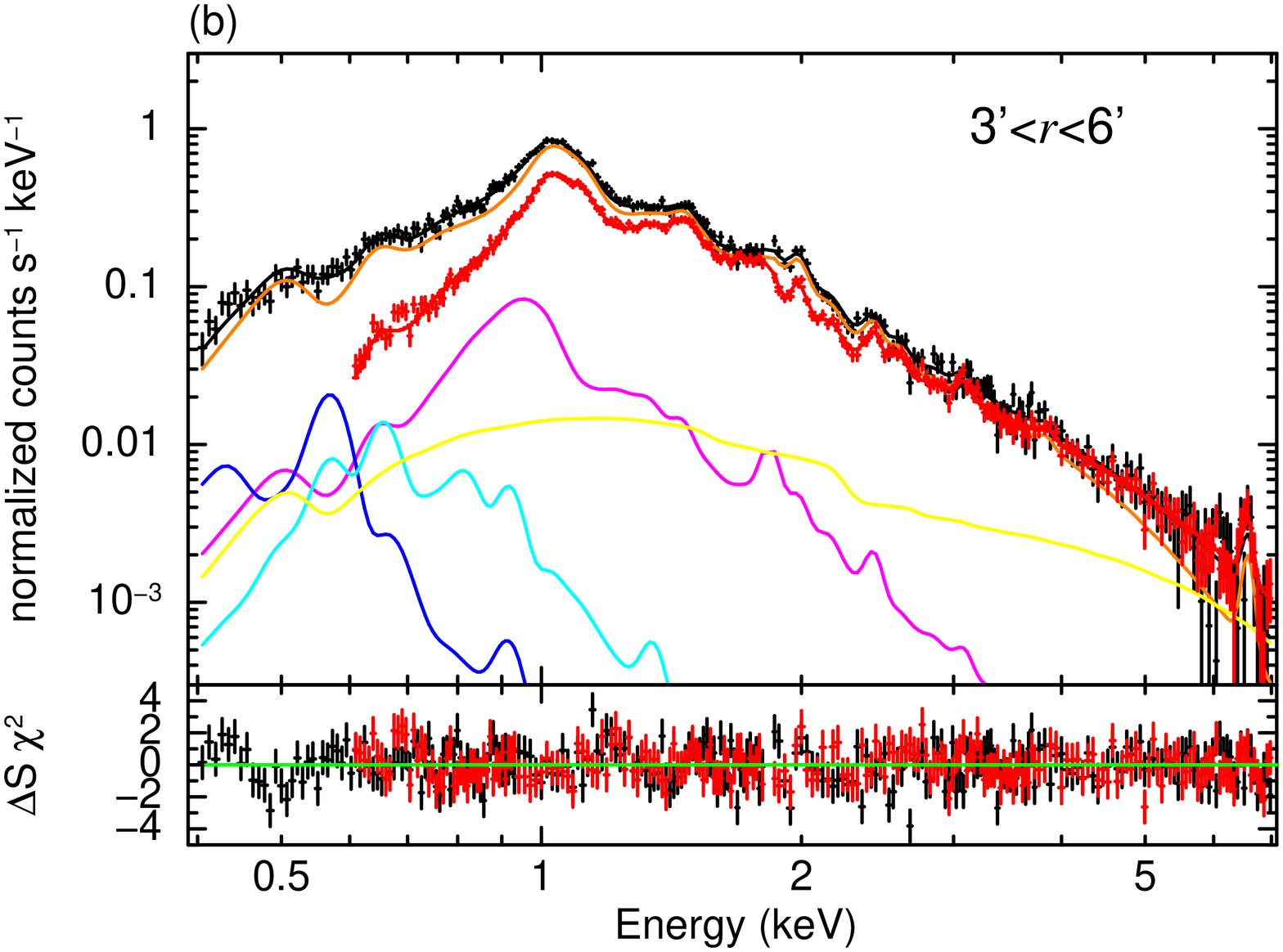}
\end{minipage}\hfill
\begin{minipage}{0.33\textwidth}
\FigureFile(\textwidth,\textwidth){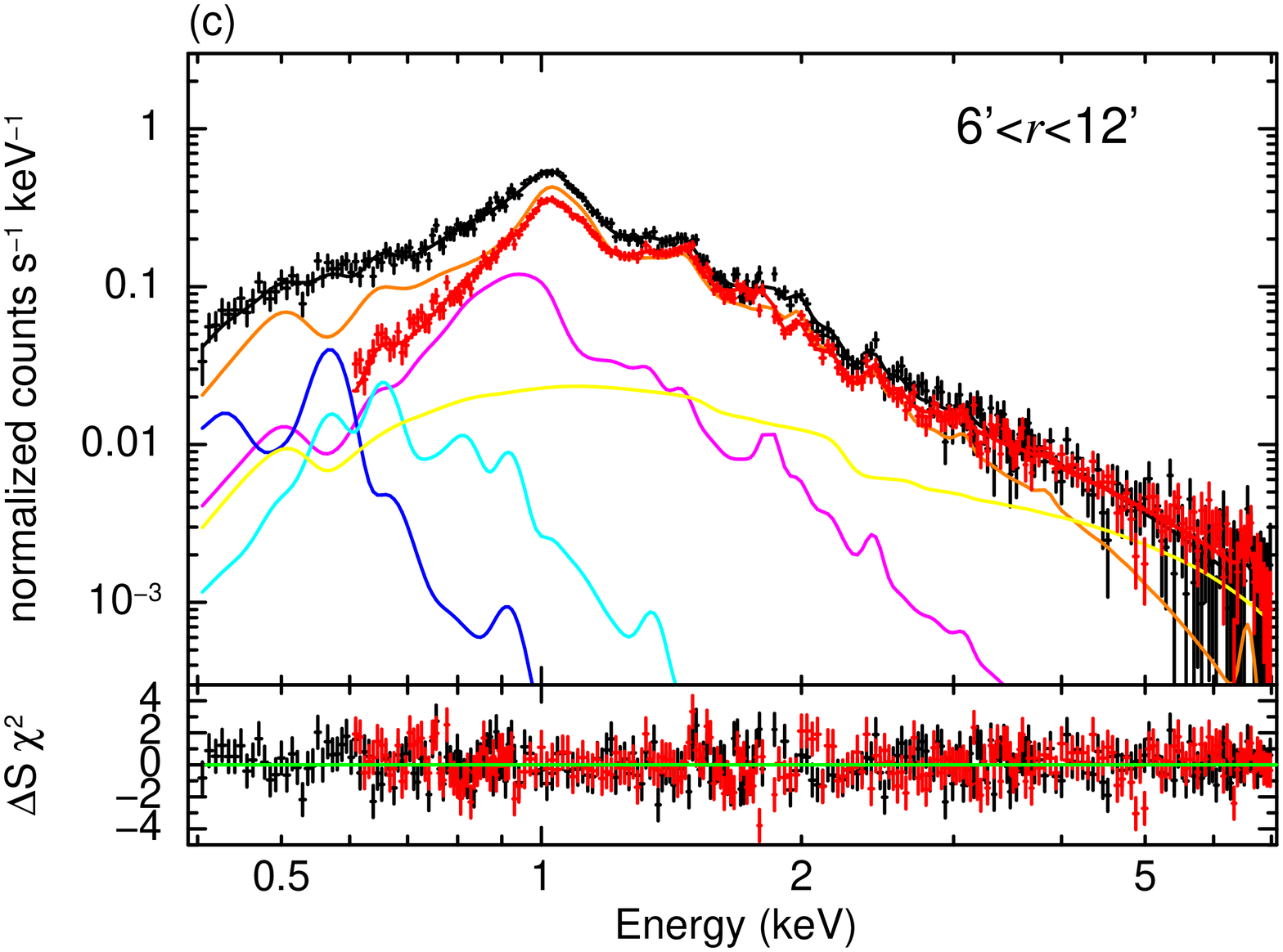}
\end{minipage}

\begin{minipage}{0.33\textwidth}
\FigureFile(\textwidth,\textwidth){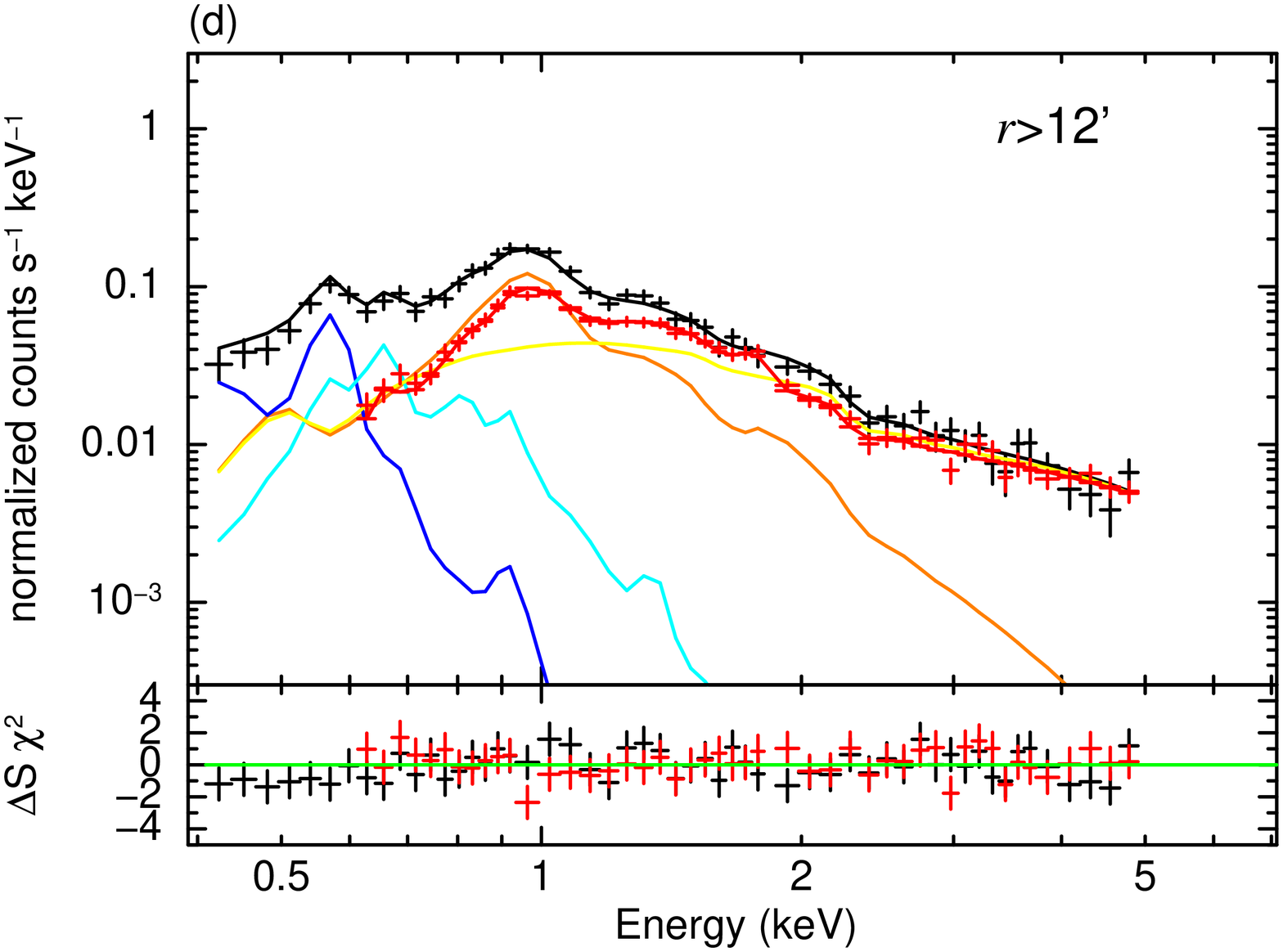}
\end{minipage}\hfill
\begin{minipage}{0.66\textwidth}
\caption{The panels show the observed spectra
for the annular regions of NGC~1550 which are denoted in the panels,
and the data are plotted by red and black crosses for BI and FI, respectively.
The estimated NXB components are subtracted,
and the black and red lines show the best-fit model 
for the BI and FI spectra, respectively.
The BI spectra of the ICM component are shown in orange and 
magenta lines. The CXB components are shown in a yellow line, 
and the Galactic emissions are indicated by blue and cyan lines, respectively.
The energy range around the Si K-edge (1.825--1.840 keV) is ignored
in the spectral fits.
The lower panels show the fit residuals in units of $\sigma$.
}\label{fig:2}
\end{minipage}
\end{figure*}

\section{Observations and Data Reduction}\label{sec:obs}

\subsection{Observations}
\label{subsec:obs}

Suzaku observed the central and offset regions of NGC~1550 in August 2008
(PI:~K. Sato).  The observation logs are given in table~\ref{tab:1}, 
and the XIS image in 0.5--2~keV is shown in figure~\ref{fig:1}.  
We analyzed only the XIS data in this paper, because the temperature 
of the HXD PIN is slightly higher during the observations.
The XIS instrument consists of three sets 
of X-ray CCDs (XIS~0, 1, and 3). XIS~1 is a back-illuminated (BI) 
sensor, while XIS~0 and 3 are front-illuminated (FI)\@.  The 
instrument was operated in the normal clocking mode (8~s exposure 
per frame), with the standard $5\times 5$ or $3\times 3$ editing mode.
During these observations, the significant effect of the 
Solar Wind Charge eXchange (SWCX) was not confirmed in ACE data
\footnote{http://www.srl.caltech.edu/ACE/ASC}, 
although it was known that the SWCX affected the Suzaku spectra 
in the lower energy range as reported in \citet{fujimoto07} and 
\citet{yoshino09}.

\subsection{Data Reduction}

We used version 2.2 processing data, and the analysis was performed 
with HEAsoft version 6.6.3 and XSPEC 12.5.0ac.
Here we give just a brief description of the data reduction.
The light curve of each sensor in the 0.3--10~keV range with a 16~s time
bin was also examined in order to exclude periods with anomalous event
rates which were greater or less than $\pm 3\sigma$ around the mean to 
remove the charge exchange contamination \citet{fujimoto07}, 
while Suzaku data was little affected by the soft proton flare compared 
with XMM data.
The exposure after the screening was essentially the same as that
before screening in table~\ref{tab:1}, which indicated that the non
X-ray background (NXB) was almost stable during the observation. 
Event screening with cut-off rigidity (COR) was not performed in our data.
In order to subtract the NXB and the extra-galactic
cosmic X-ray background (CXB), we employed the dark Earth
database by the ``xisnxbgen'' Ftools task.

We generated two Ancillary Response Files (ARFs) for the spectrum 
of each annular sky region, $A^{\makebox{\small\sc u}}$ and 
$A^{\makebox{\small\sc b}}$, which respectively assumed uniform sky 
emission and $\sim 1^{\circ} \times 1^{\circ}$ size of the $\beta$-model 
surface brightness profile, $\beta = 0.47$ and $r_c =$ \timeform{0.85'}, 
in \citet{fukazawa04}, by the ``xissimarfgen'' Ftools task \citep{ishisaki07}.
We also included the effect of the contaminations on the optical blocking 
filter of the XISs in the ARFs.
Since the energy resolution also slowly degraded
after the launch, due to radiation damage, this effect was included
in the Redistribution Matrix File (RMF) by the ``xisrmfgen'' Ftools task.

\section{Temperature and Abundance Profiles}

\begin{figure*}
\begin{minipage}{0.33\textwidth}
\FigureFile(\textwidth,\textwidth){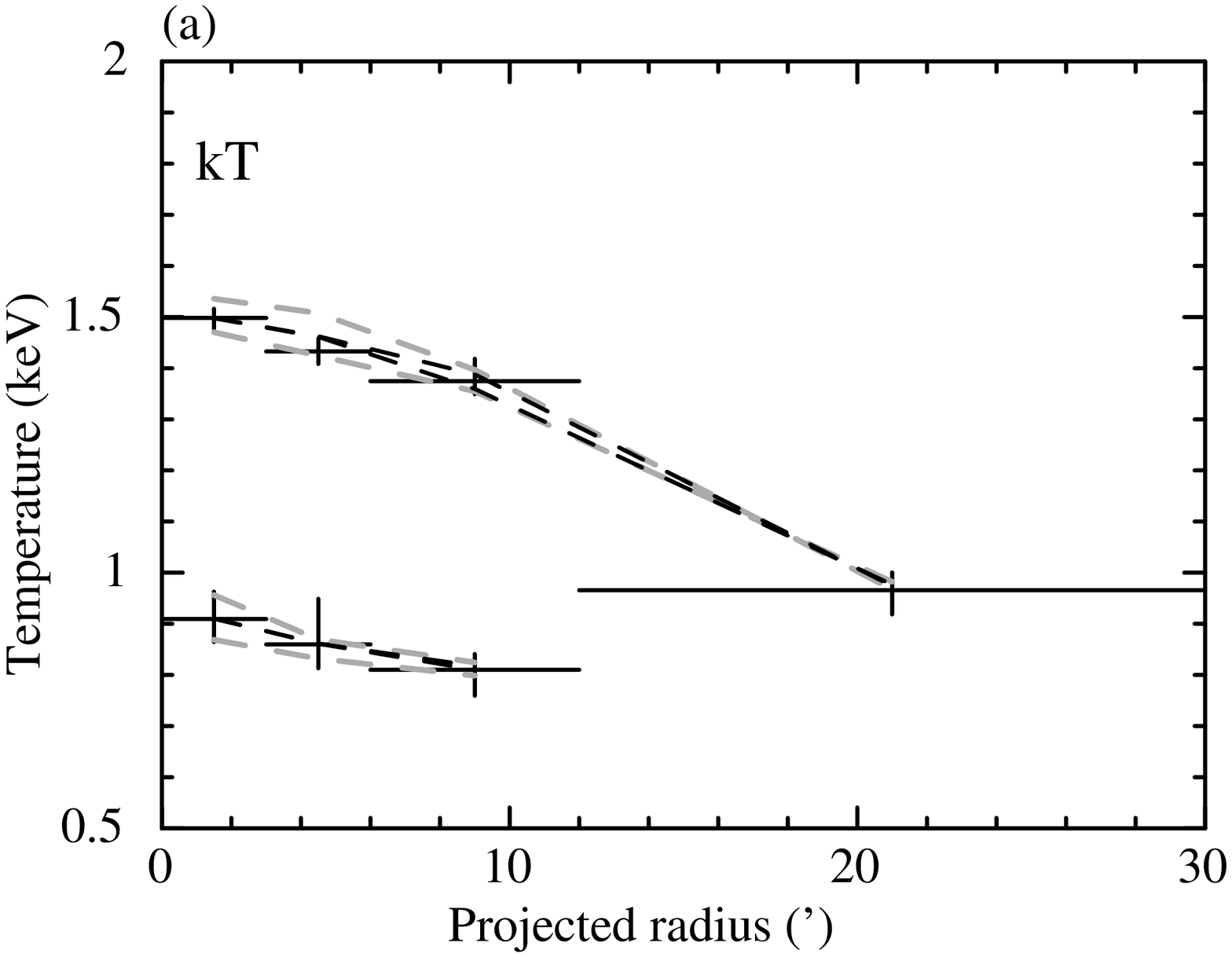}
\end{minipage}\hfill
\begin{minipage}{0.33\textwidth}
\FigureFile(\textwidth,\textwidth){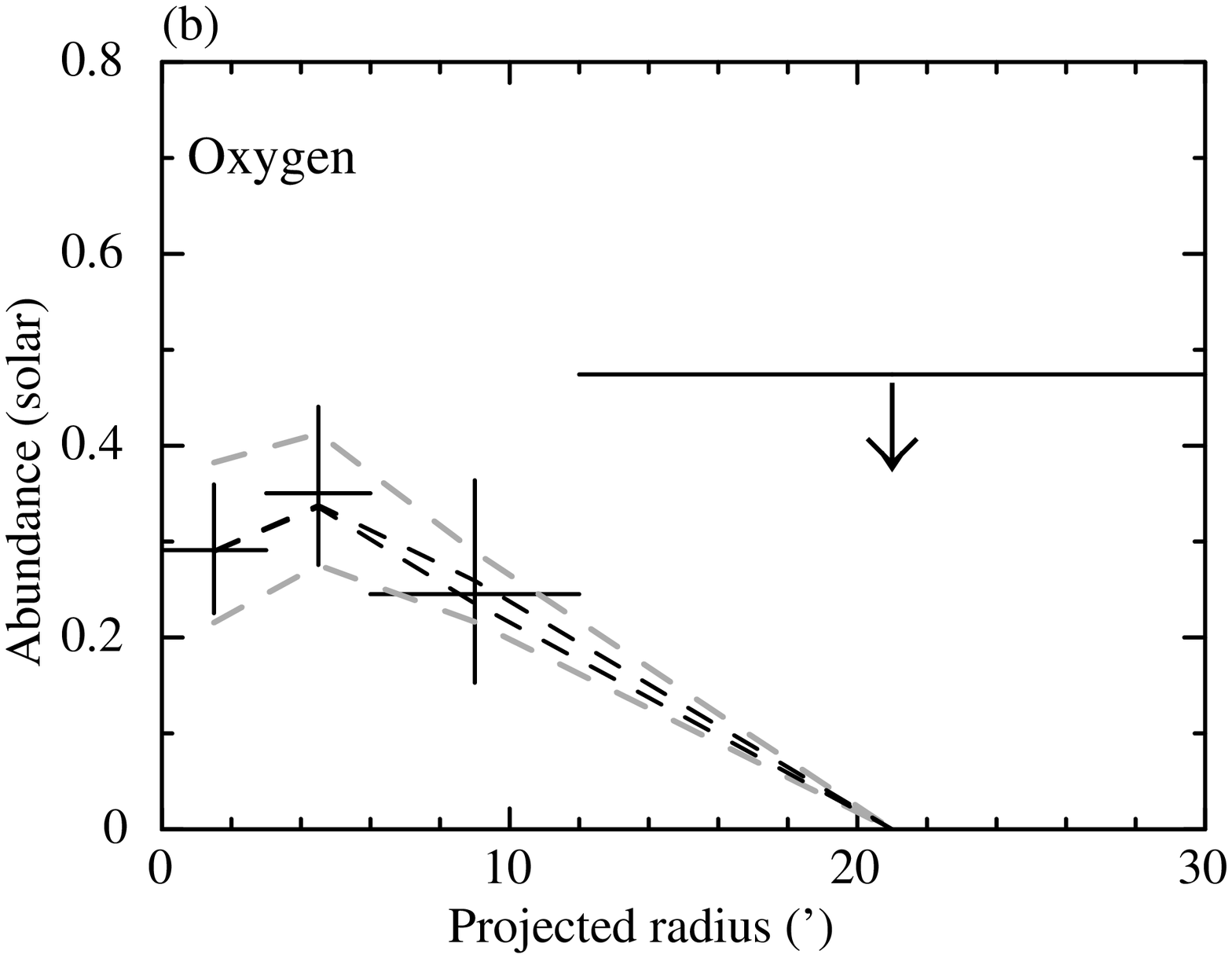}
\end{minipage}\hfill
\begin{minipage}{0.33\textwidth}
\FigureFile(\textwidth,\textwidth){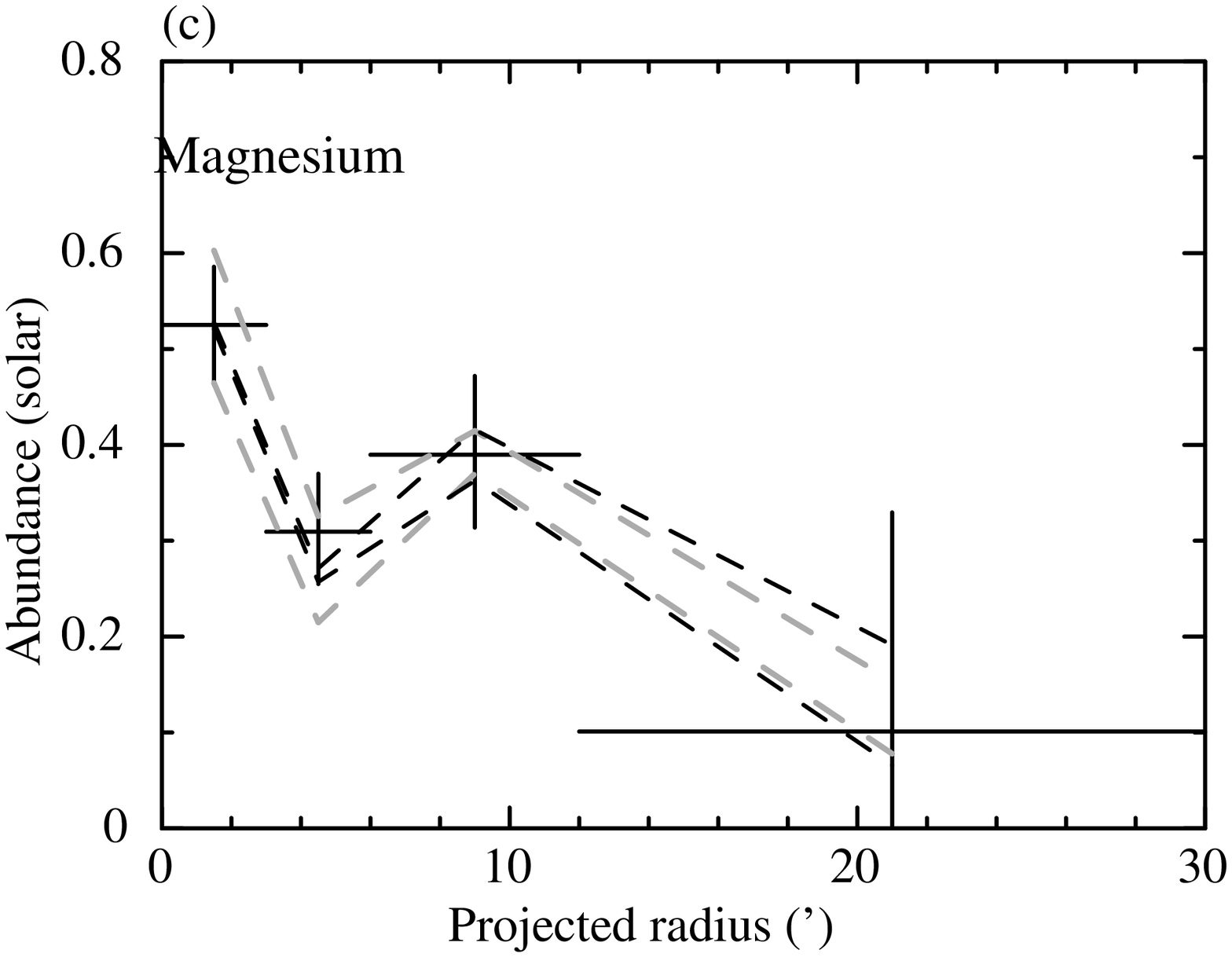}
\end{minipage}

\begin{minipage}{0.33\textwidth}
\FigureFile(\textwidth,\textwidth){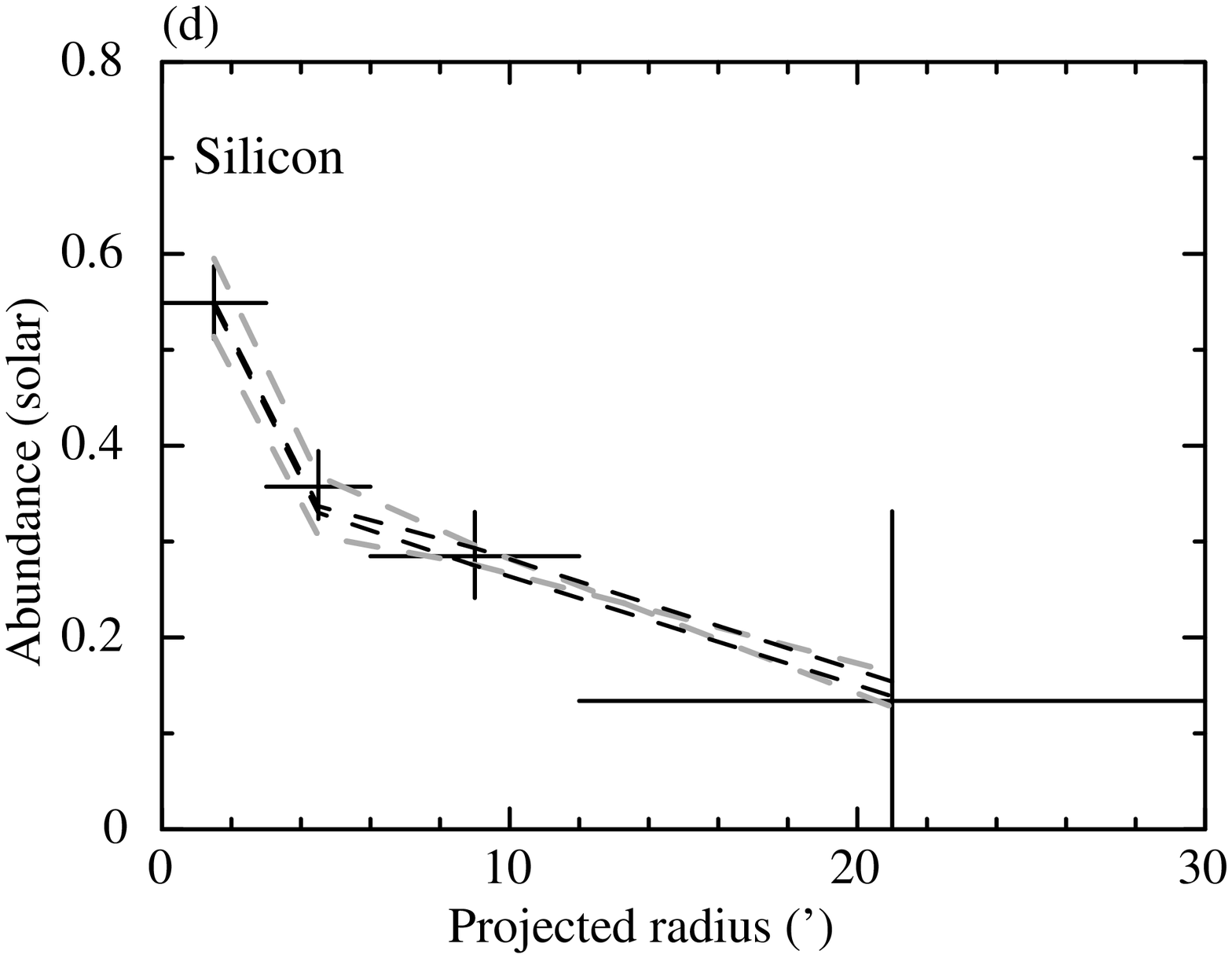}
\end{minipage}\hfill
\begin{minipage}{0.33\textwidth}
\FigureFile(\textwidth,\textwidth){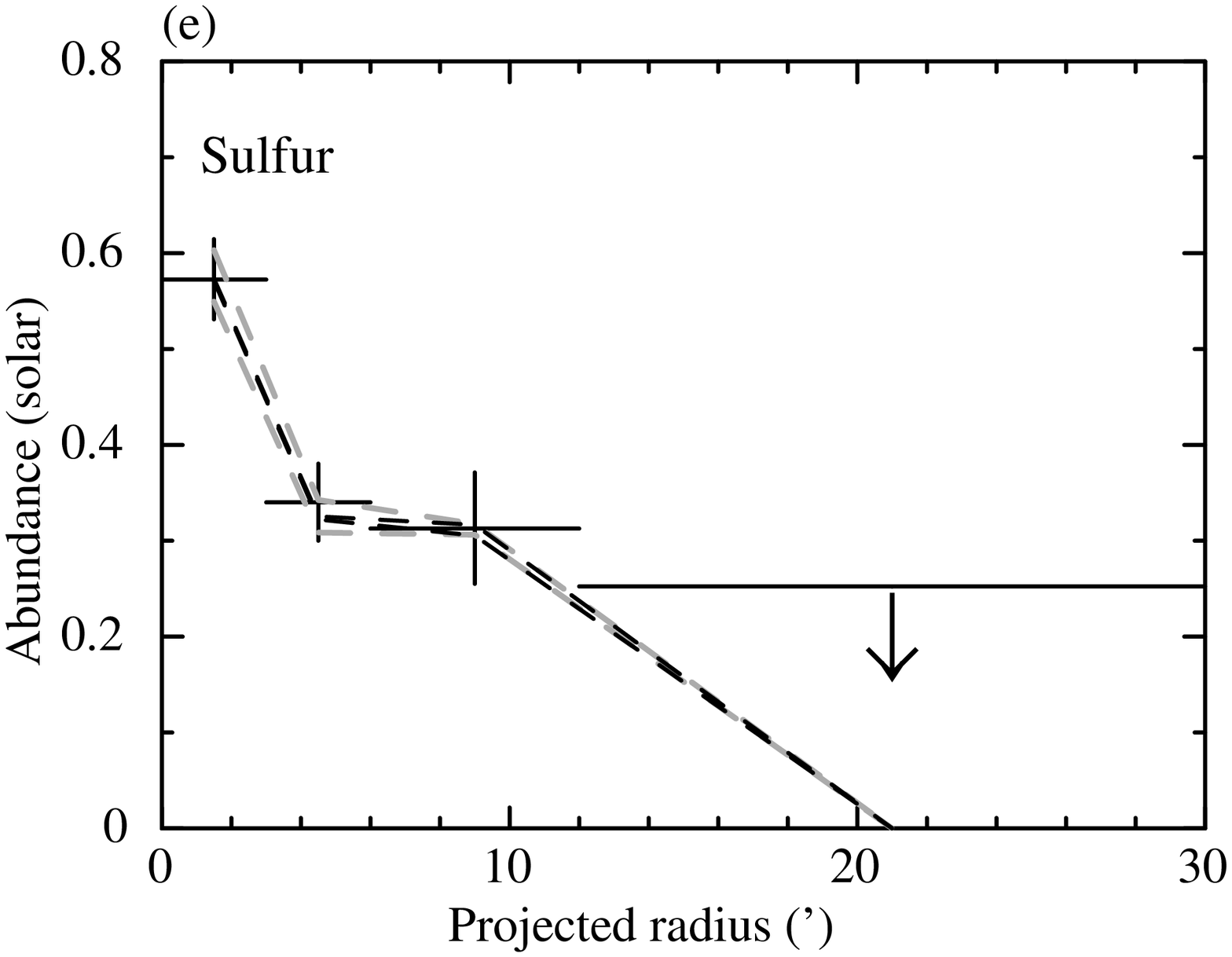}
\end{minipage}\hfill
\begin{minipage}{0.33\textwidth}
\FigureFile(\textwidth,\textwidth){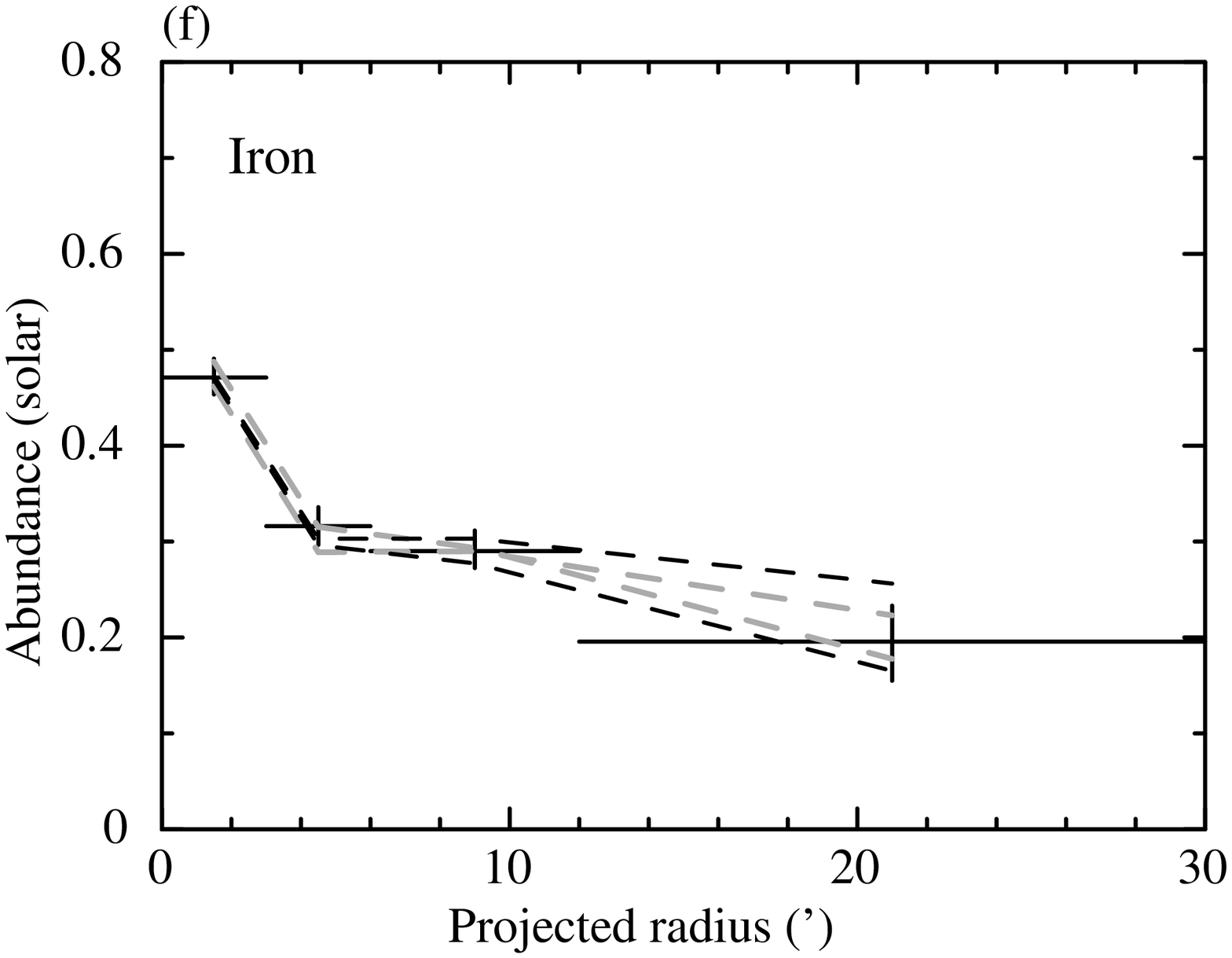}
\end{minipage}

\vspace*{-0.5ex}
\caption{(a): Radial temperature profiles derived from the spectral
fits for each annulus against the projected radius.  
Black dashed lines show systematic change of the best-fit
values by varying the thickness of the OBF contaminant by $\pm 10$\%.  
Light-gray dashed lines denote those
when the NXB levels are varied by $\pm 10$\%.
(b)--(f): Radial abundance profiles are plotted in the same
way as in (a).}
\label{fig:3}
\end{figure*}

\subsection{Spectral Fit}
\label{sec:spec}

We extracted spectra from five annular regions
of 0$'$--3$'$, 3$'$--6$'$, and 6$'$--12$'$ for the central observation, 
and whole area, which corresponds to 12$'$--30$'$, for the offset region,
centered on (RA, Dec) = (\timeform{4h19m37.9s}, \timeform{+02D24'36''}).
Each annular spectrum is shown in figure~\ref{fig:2}.
The ionized Mg, Si, S, Fe lines are clearly seen in each region.
The O\emissiontype{VII} and O\emissiontype{VIII} lines were prominent
in the outer rings, however, most of the O\emissiontype{VII} line
was considered to come from the local Galactic emission,
and we dealt with those in the same way 
as mentioned in \citet{sato07a,sato08,sato09a,sato09b}.

The spectra with BI and FI for all regions were fitted simultaneously
in the energy range, 0.4--7.1/0.4--5.0 keV (BI) and 
0.6--7.1/0.6--5.0 (FI) for the central/offset observations, 
respectively.  In the simultaneous fit, the
common Galactic emission and CXB components were included 
for all regions.  We excluded the narrow energy band around the Si K-edge
(1.825--1.840~keV) because its response was not modeled correctly.
The energy range below 0.4~keV was also excluded because the C edge
(0.284~keV) seen in the BI spectra could not be reproduced well in our
data.  The range above 7.1~keV was also ignored because Ni line ($\sim
7.5$~keV) in the background left a spurious feature after the NXB
subtraction at large radii.  In the simultaneous fits of BI and FI
data, only the normalization parameter was allowed to take different
values between them.

It is important to estimate the Galactic component precisely, because 
the Galactic component gives significant contribution especially in the
outer regions, as shown in figure~\ref{fig:2}.  However, the ICM
component is still dominant in almost all the energy range except for
the O\emissiontype{VII} line.  We assumed two temperature {\it apec}
model (assuming 1 solar abundance and zero redshift) for the
Galactic component, and fitted the data with the following model
formula: ${\it constant}\times( {\it apec}_{\rm cool}+
{\it phabs} \times ({\it apec}_{\rm
hot} + {\rm ICM} ))$ as shown in \citet{yoshino09}.
As a result, the best-fit temperatures of 0.098 and 0.219 keV 
for the Galactic models as shown in table \ref{tab:2} 
are consistent with the values as shown 
in \citet{yoshino09} and \citet{lumb02}.
Thus, we concluded that the two temperature model of 
the Galactic emission was enough to represent NGC~1550 data. 
The resultant normalizations of the {\it apec} models in table~\ref{tab:2}
are scaled so that they give the surface brightness in unit solid
angle of arcmin$^2$, and are constrained to give the same surface
brightness and the same temperature for the simultaneous fits of all
annuli\footnote{The Normalizations of ``{\it norm}$_1$'' in 
table 5 in \citet{sato08} are an order of magnitude larger than 
the actual values.}.

\begin{table*}
\caption{Summary of the parameters of the fits to
each annular spectrum of NGC~1550.  All annuli were simultaneously
fitted.
Errors are 90\% confidence range of statistical errors, and
do not include systematic errors.  The solar abundance ratio of {\it
angr} was assumed.  These results are plotted in figure~\ref{fig:3}.
}\label{tab:2}
\begin{center}
\begin{tabular}{lcccccc}
\hline\hline
\makebox[2em][l]{Galactic} & {\it Norm}$_1$\makebox[0in][l]{\,$^\ast$} & $kT_1$& {\it Norm}$_2$\makebox[0in][l]{\,$^\ast$} & $kT_2$&  &  \\
& &(keV)&&(keV)& & \\
\hline
     & 3.37$^{+1.43}_{-1.14}$ & 0.098$^{+0.005}_{-0.014}$ & 0.92$^{+0.23}_{-0.58}$ & 0.219$^{+0.062}_{-0.046}$ &  &  \\
\hline\hline
\makebox[2em][l]{ICM} & {\it Norm}$_1$\makebox[0in][l]{\,$^\dagger$} & $kT_1$& {\it Norm}$_2$\makebox[0in][l]{\,$^\dagger$} & $kT_2$& {\it Norm}$_1$/{\it Norm}$_2$ & $\chi^2$/dof\makebox[0in][l]{\,$^\ddagger$} \\
& &(keV)&&(keV)& & \\
\hline
0$'$--3$'$    &  358.5$^{+11.0}_{-6.1}$  & 1.50$^{+0.02}_{-0.02}$ &  58.8$^{+4.8}_{-4.7}$  & 0.91$^{+0.05}_{-0.05}$ & 6.10$^{+0.53}_{-0.49}$ &  613/463\\
3$'$--6$'$    &  88.2$^{+3.2}_{-1.8}$  & 1.43$^{+0.03}_{-0.02}$ & 5.4$^{+1.3}_{-0.9}$ & 0.86$^{+0.09}_{-0.05}$ & 16.48$^{+3.94}_{-2.89}$ & 531/469\\
6$'$--12$'$    & 28.3$^{+1.6}_{-0.9}$  & 1.37$^{+0.04}_{-0.03}$ & 5.2$^{+0.5}_{-2.8}$ & 0.81$^{+0.03}_{-0.05}$ &  5.42$^{+0.58}_{-2.96}$ & 520/469\\
12$'$--30$'$    &  5.0$^{+1.1}_{-0.6}$  & 0.97$^{+0.03}_{-0.05}$ & -- & -- & -- & 73/91\\
total  &  &  &  &  &  & 1737/1492\\\\[-2ex]
\hline\hline
\makebox[2em][l]{ICM} &O&Ne&Mg,Al&Si&\makebox[0in][c]{S,Ar,Ca}&Fe,Ni \\
&(solar)&(solar)&(solar)&(solar)&(solar)&(solar) \\
\hline
0$'$--3$'$    & 0.29$^{+0.07}_{-0.07}$ & 0.50$^{+0.12}_{-0.12}$ & 0.53$^{+0.06}_{-0.06}$ & 0.55$^{+0.04}_{-0.04}$ & 0.57$^{+0.04}_{-0.04}$ & 0.47$^{+0.02}_{-0.02}$ \\
3$'$--6$'$    & 0.35$^{+0.09}_{-0.07}$ & 0.26$^{+0.11}_{-0.10}$ & 0.31$^{+0.06}_{-0.05}$ & 0.36$^{+0.04}_{-0.03}$ & 0.34$^{+0.04}_{-0.04}$ & 0.32$^{+0.02}_{-0.02}$ \\
6$'$--9$'$    & 0.25$^{+0.12}_{-0.09}$ & 0.39$^{+0.14}_{-0.14}$ & 0.39$^{+0.08}_{-0.08}$ & 0.28$^{+0.05}_{-0.04}$  & 0.31$^{+0.06}_{-0.06}$ & 0.29$^{+0.02}_{-0.02}$ \\
9$'$--17$'$    & 0.00$^{+0.47}_{-0.00}$ & 0.01$^{+0.38}_{-0.01}$ & 0.10$^{+0.23}_{-0.10}$ & 0.13$^{+0.20}_{-0.13}$ & 0.00$^{+0.25}_{-0.00}$ & 0.20$^{+0.04}_{-0.04}$ \\
\hline\\[-1ex]
\multicolumn{7}{l}{\parbox{0.8\textwidth}{\footnotesize 
\footnotemark[$\ast$] 
Normalization of the {\it apec} component
divided by the solid angle, $\Omega^{\makebox{\tiny\sc u}}$,
assumed in the uniform-sky ARF calculation (20$'$ radius),
${\it Norm} = \int n_{\rm e} n_{\rm H} dV \,/\,
(4\pi\, (1+z)^2 D_{\rm A}^{\,2}) \,/\, \Omega^{\makebox{\tiny\sc u}}$
$\times 10^{-20}$ cm$^{-5}$~arcmin$^{-2}$, 
where $D_{\rm A}$ is the angular distance to the source.}}\\
\multicolumn{7}{l}{\parbox{0.8\textwidth}{\footnotesize
\footnotemark[$\dagger$] 
Normalization of the {\it vapec} component scaled with a factor of
the selected region comparing to the assumed image in
 ``xissimarfgen'',
${\it Norm}= {factor} \int
n_{\rm e} n_{\rm H} dV \,/\, [4\pi\, (1+z)^2 D_{\rm A}^{\,2}]$ $\times
10^{-20}$~cm$^{-5}$~arcmin$^{-2}$, where $D_{\rm A}$ is the angular
distance to the source. }}\\
\multicolumn{7}{l}{\parbox{0.8\textwidth}{\footnotesize
\footnotemark[$\ddagger$] 
All regions were fitted simultaneously.
}}
\end{tabular}
\end{center}
\end{table*}

\begin{table*}
\caption{List of $\chi^2$/dof for the fits of the nominal and
considering the systematic errors such as contaminant of OBF and 
background level. For details, see text.}
\label{tab:3}
\begin{center}
\begin{tabular}{lccccc}
\hline\hline
\makebox[6em][l]{Region} & nominal &\multicolumn{2}{c}{contaminant} & \multicolumn{2}{c}{background}\\
\hline
 & & \makebox[0in][c]{+10\%} & \makebox[0in][c]{-10\%} & \makebox[0in][c]{+10\%} & \makebox[0in][c]{-10\%}\\
\hline
All $\dotfill$   & 1737/1492 & 1741/1492 & 1803/1492 & 1745/1492 & 1747/1492 \\
\hline
\end{tabular}
\end{center}
\end{table*}

The ICM spectra for the central observation, $r<12'$, were clearly 
better represented by two {\it vapec} models than one {\it vapec} model in
the $\chi^2$ test. On the other hand, the ICM spectra for the offset
observation, $r=12$--$30'$, were well-presented by a single temperature model.
Thus, we carried out the simultaneous fit with the following formula
of the Galactic and ICM components: ${\it constant}\times( {\it
apec}_{\rm cool}+{\it phabs} \times ( {\it apec}_{\rm hot} +
{\it vapec}_{0<r<30'} + {\it vapec}_{0<r<12'})$\@.  The fit results are
shown in table \ref{tab:2}.  The abundances were linked in the
following way, Mg=Al, S=Ar=Ca, Fe=Ni, which gave good constraint
especially for the offset regions.  The abundances were also linked
between the two ${\it vapec}$ components for $r<12'$ region.  Results
of the spectral fit for individual annuli are summarized in
table~\ref{tab:2} and figure~\ref{fig:3}, in which systematic error
due to the OBF contamination and NXB estimation are shown.

\subsection{Temperature Profile}
\label{subsec:radial}

Radial temperature profile and the ratio of the {\it vapec}
normalizations between the hot and cool ICM components are shown in
figure~\ref{fig:3}(a) and table~\ref{tab:2}. 
The ICM temperature of hot and cool components at the central region 
was $\sim1.5$ and $\sim0.9$ keV, respectively, and the temperature 
decreased mildly to $\sim1.0$ keV in the outermost region, while the 
cool components were almost constant at $\sim0.9$ keV.
Our results for the two temperature ICM model are consistent with the
XMM result \citep{kawaharada09}.
For the hot component, our results are also consistent with 
the previous Chandra result \citep{sun03}. 
The radius of $30'\sim 457$~kpc corresponds to $\sim 0.47\; r_{\rm
180}$, and the temperature decline is clearly recognized 
in such a small system out to this radius.

\subsection{Abundance Profiles}

Metal abundances are determined for the six element groups
individually as shown in figures~\ref{fig:3}(b)--(f).  The four
abundance values for Mg, Si, S, and Fe and their radial variation look
similar to each other.  The central abundances lie around 
$\sim 0.5$--0.6 solar, and they commonly decline to about 1/4 of the central
value in the outermost annulus.  On the other hand,
the O profile looks flatter compared with the other elements.  
Because the results for the offset regions had large errors, we examined the
summed spectra for these regions.  
We noted that, when we examined all abundances to be free in the fits,
the resultant parameters did not change within the statistical errors. 
In addition, even if all regions were fitted by a two temperature model,
the resultant abundance profiles did not change within the statistical errors.

We also examined the systematic error of our results by changing the
background normalization by $\pm 10$\%, and the error range is plotted
with light-gray dashed lines in figure~\ref{fig:3}. The systematic
error due to the background estimation is almost negligible.  The
other systematic error concerning the uncertainty in the OBF
contaminant is shown by black dashed lines as shown in
figure~\ref{fig:3}.  A list of $\chi^2$/dof by changing the systematic
errors is presented in table \ref{tab:3}.  
We note that Ne abundance is not
reliably determined due to an overlap with the strong and complex Fe-L
line emissions, however we left these abundance to vary freely during
the spectral fit.

\section{Discussion}\label{sec:discuss}

\subsection{Metallicity Distribution in the ICM}
\label{subsec:metal}

\begin{figure}
\centerline{\FigureFile(0.45\textwidth,8cm){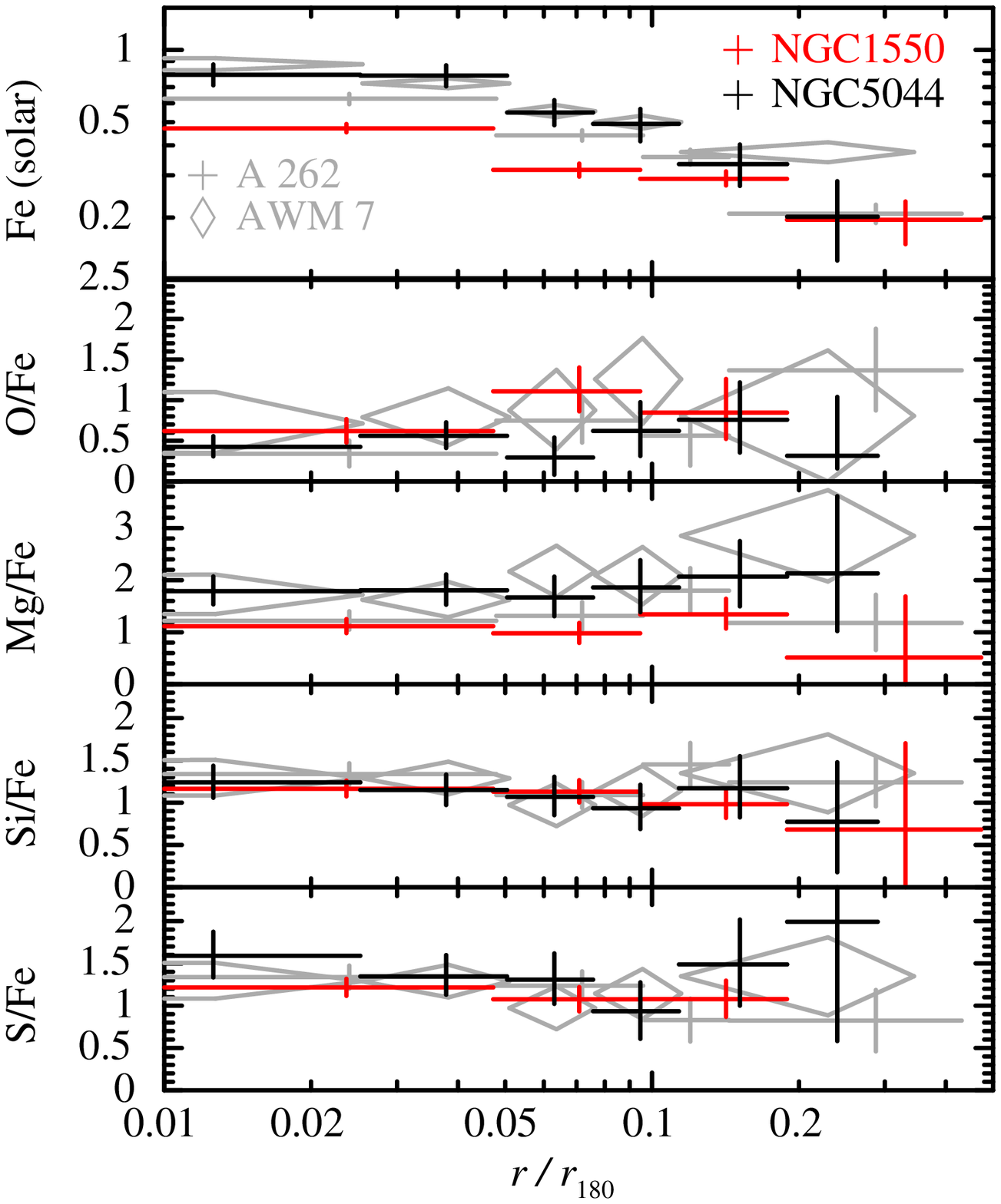}}
\vspace*{-1ex}
\caption{ Comparison of the Fe abundance and the other metals to the Fe 
abundance ratios for NGC~1550 (red crosses) with
those for the NGC~5044 group (black crosses; \cite{komiyama09}), 
A~262 cluster (light gray crosses; \cite{sato09b}) 
and AWM~7 cluster (light gray diamonds; \cite{sato08}). 
The Fe abundance and the O/Fe,
Mg/Fe, Si/Fe, and S/Fe abundance ratios in solar units
\citep{anders89} are plotted against the projected radius scaled by
the virial radius, $r_{180}$, in all panels.  }\label{fig:4}
\end{figure}

\begin{figure}
\centerline{\FigureFile(0.45\textwidth,8cm){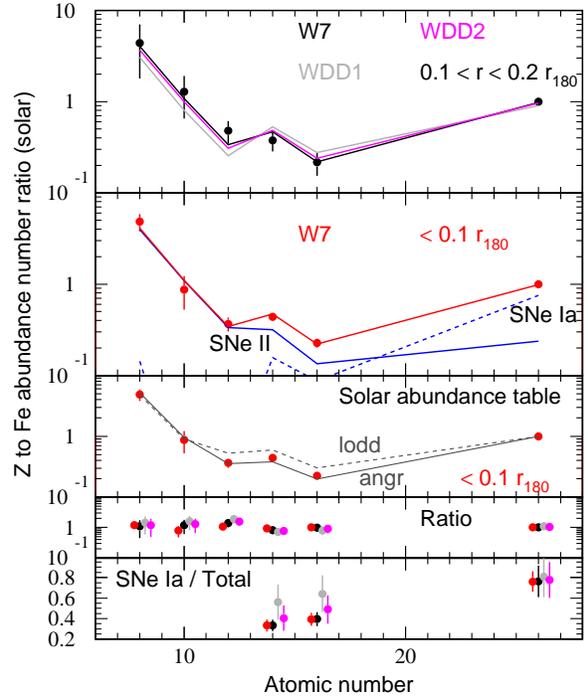}}
\vspace*{-1ex}
\caption{
Fit results of number ratios of elements to Fe of NGC~1550.
Top and second panel show the abundance number ratios 
in solar unit in $0.1<r<0.2\;r_{\rm 180}$ (black) 
and $0.1\;r_{\rm 180}$ (red) regions, respectively.
Blue dashed and solid lines in the second panel correspond to 
the contributions of SNe~Ia (W7) and SNe~II within $0.1\;r_{180}$, 
respectively.
Ne (atomic number = 10) is excluded in the fit.
Third panel shows the comparison with solar abundance table of 
\citet{anders89} (light-gray line) and \citet{lodders03} (dashed
 light-gray line). 
Forth panel indicates ratios of data points to the best-fit model.
Bottom panel indicates fractions of the SNe~Ia contribution 
to total metals in the best-fit model for each element, respectively.
}\label{fig:5}
\end{figure}

Suzaku observation of NGC~1550 confirmed that the metals in the ICM of
this fossil group are indeed extending to a large radius.  The
measured elements are O, Mg, Si, S, and Fe out to a radius of
$30'\simeq 457$~kpc, which corresponds to $\sim0.47~r_{180}$, as shown
in figure \ref{fig:3}. The Ne abundance had a large uncertainty due to
the strong coupling with Fe-L lines.  Distributions of Mg, Si, S, and
Fe are similar to each other, while O profile shows no central peak
and large error in the outer region at $r > 12'$.  We plotted
abundance ratios of O, Mg, Si, and S to Fe as a function of the
projected radius in figure~\ref{fig:4}.  The ratios Mg/Fe, Si/Fe and
S/Fe are consistent to be a constant value around 1.5--2, while O/Fe
ratio in the innermost region ($r<3'$) is significantly lower around
0.5\@.  In addition, the O/Fe ratio suggests some increase with
radius.  Note that these abundance profiles are not deconvolved and
are averaged over the line of sight.

Recent Suzaku observations have presented abundance profiles in
several relaxed cooling core groups and clusters: a group of galaxies
NGC~5044 \citep{komiyama09}, a poor cluster of galaxies Abell~262
\citep{sato09b}, and AWM~7 \citep{sato08}.  We compare Fe abundance
and O, Mg, Si, S to Fe abundance ratios of NGC~1550 with those of
NGC~5044, Abell~262 and AWM~7 as shown in figure~\ref{fig:4}.  While
NGC~1550 shows slightly lower Fe abundance than those of NGC~5044,
Abell~262, and AWM~7 in the central region ($r\lesssim 0.1~r_{180}$), the
abundance in the outer region ($r\gtrsim0.3~r_{180}$) are quite
similar, showing a decrease to $\sim0.2$ solar.  On the other hand,
the abundance ratios of O/Fe, Mg/Fe, Si/Fe, and S/Fe are quite similar
between the four systems.

\citet{matsushita10} reports the Fe radial profiles of 28 clusters 
of galaxies observed with XMM. The Fe abundance profile of NGC~1550 
also has similar feature to those of clusters up to $\sim0.5~r_{180}$.
Although \citet{rasmussen09} suggests that the radial Si profiles in group 
have softer than the Fe profiles, our results show the constant values 
of the Si to Fe ratios up to $\sim0.5~r_{180}$ as shown in figure 
\ref{fig:4}. 

\subsection{Number Ratio of SNe~II to SNe~Ia}

\begin{table*}
\caption{
Integrated number of SNe~I ($N_{\rm Ia}$) and
number ratio of SNe~II to SNe~Ia ($N_{\rm II}$/$N_{\rm Ia}$).
}\label{tab:4}
\begin{center}
\begin{tabular}{lcccr}
\hline
\multicolumn{1}{c}{Region} &
\makebox[0in][c]{SNe~Ia Model} &
$N_{\rm Ia}$& $N_{\rm II}$/$N_{\rm Ia}$ & $\chi^2$/dof \\
\hline
 $<0.1\;r_{180}$ &W7 & $1.3\pm0.1\times10^{8}$& $2.9\pm 0.5$ & 3.7/3\\
 0.1 -- 0.2~$r_{\rm 180}$ &W7 & $2.6\pm0.4\times10^{8}$& $2.8\pm 1.0$ & 5.7/3\\
 $< 0.1\;r_{\rm 180}$ &WDD1 & $1.3\pm0.1\times10^{8}$& $2.1\pm 0.5$ & 44.2/3\\
 0.1 -- 0.2~$r_{\rm 180}$ &WDD1 & $2.8\pm0.5\times10^{8}$& $2.0\pm 1.0$ & 21.0/3\\
 $< 0.1\;r_{\rm 180}$ &WDD2 & $1.2\pm0.1\times10^{8}$& $2.8\pm 0.6$ & 10.4/3\\
 0.1 -- 0.2~$r_{\rm 180}$ &WDD2 & $2.4\pm0.4\times10^{8}$& $2.8\pm 1.1$ & 8.9/3\\
\hline
\end{tabular}
\end{center}
\end{table*}

In order to examine relative contributions from SNe~Ia and SNe~II to
the ICM metals, the elemental abundance pattern of O, Mg, Si, S and Fe
was examined for the inner ($r<0.1~r_{180}$) and the immediate outer
(0.1--$0.2~r_{180}$) regions.  The abundance ratios to Fe were fitted
by a combination of average SNe~Ia and SNe~II yields per supernova, as
shown in figure~\ref{fig:5}.  The fit parameters were the integrated
number of SNe~Ia ($N_{\rm Ia}$) and the number ratio of SNe~II to
SNe~Ia ($N_{\rm II}/N_{\rm Ia}$), because $N_{\rm Ia}$ could be well
constrained by the relatively small errors in the Fe abundance.  The
SNe Ia and II yields were taken from \citet{iwamoto99} and
\citet{nomoto06}, respectively.  We assumed the Salpeter IMF for
stellar masses from 10 to 50 $M_{\odot}$ with the progenitor
metallicity of $Z=0.02$ for SNe~II, and W7, WDD1 or WDD2 models for
SNe~Ia.  Table~\ref{tab:4} and figure~\ref{fig:5} summarize the fit
results.  The number ratios were better represented by the W7 SNe~Ia
yield model than by WDD1\@. The number ratio of SNe~II to SNe~Ia with
W7 is $\sim2.9$ within 0.1~$r_{180}$, while the ratio assuming WDD1 is
$\sim 2.1$.  The WDD2 model gave the result very similar to the W7
value.  The resultant number ratios are consistent with the previous
result by \citet{sato07b}\@.  We also compared the abundance pattern
of NGC~1550 with the solar abundance.  The third panel in
figure~\ref{fig:5} shows this comparison for $r<0.1~r_{180}$ of
NGC~1550 with two different solar abundance patterns given by
\citet{anders89} and \citet{lodders03}.  Abundances of Mg, Si, and S
fall between the two solar abundance patterns.

Almost $\sim80$\% of Fe and $\sim40$\% of Si and S were synthesized by
SNe Ia in the W7 model, as demonstrated in the bottom panel of
figure~\ref{fig:5}. These observed features of the fossil group are
similar to those for clusters with $kT =2-4$ keV clusters studied by
\citet{sato07b} and \citet{sato09b}\@.  The values in
table~\ref{tab:4} imply that the $N_{\rm II}/N_{\rm Ia}$ ratio for the
inner and outer regions behave in the similar manner for different
supernova models. We note that the fit was not formally acceptable
based on the $\chi^2$ value in table~\ref{tab:4}\@. As described in
\citet{sato07b}, the models adapted here (SNe yield, Salpeter IMF,
etc.) are probably too simplified.

\begin{table*}
\caption{Comparison of IMLR, OMLR and MMLR with B-band luminosity 
for all systems.
}\label{tab:5}
\begin{center}
\begin{tabular}{lrrcrrrl}
\hline\hline
& IMLR & OMLR & MMLR & \multicolumn{2}{c}{$r$} & \multicolumn{1}{c}{$k\langle T \rangle$} & Reference \\
& ($M_{\odot}/L^{\rm B}_{\odot}$) & ($M_{\odot}/L^{\rm B}_{\odot}$) &
	     ($M_{\odot}/L^{\rm B}_{\odot}$) & (kpc) & ($r_{180}$) & (keV) & \\
\hline
Suzaku & & & & & & & \\
NGC~5044 $\dotfill$  & $2.6^{+0.2}_{-0.2}\times 10^{-3}$ & $6.6^{+1.9}_{-1.7}\times 10^{-3}$
	 &$1.6^{+0.2}_{-0.2}\times 10^{-3}$  &88 & 0.10 & $\sim 1.0$ &       \citet{komiyama09}\\
                    & $3.6^{+0.4}_{-0.3}\times 10^{-3}$ & $9.4^{+5.2}_{-2.1}\times 10^{-3}$
 &$2.6^{+0.4}_{-0.3}\times 10^{-3}$  &260 & 0.30 &  &    \\
NGC~1550 $\dotfill$  & $2.6^{+0.1}_{-0.1}\times 10^{-3}$ & $1.2^{+0.3}_{-0.2}\times 10^{-3}$
	 &$9.4^{+1.2}_{-1.1}\times 10^{-4}$  &46 & 0.09 & $\sim 1.2$ &
 This work\\
                    & $7.4^{+0.4}_{-0.4}\times 10^{-3}$ & $3.3^{+1.1}_{-0.8}\times 10^{-2}$
	 &$3.3^{+0.5}_{-0.5}\times 10^{-3}$  &183 & 0.19 &  &   \\
Fornax $\dotfill$  & $4\times 10^{-4}$ & $2\times 10^{-3}$ & -- &130 & 0.13 & $\sim 1.3$ &         \citet{matsushita07b} \\
NGC~507 $\dotfill$  & $6.0^{+0.4}_{-0.3}\times 10^{-4}$ & $2.6^{+0.6}_{-0.5}\times 10^{-3}$
	 &$3.7^{+0.4}_{-0.4}\times 10^{-4}$  &120 & 0.11 & $\sim 1.5$ &  \citet{sato09a}\\
                    & $1.7^{+0.2}_{-0.2}\times 10^{-3}$ & $6.6^{+3.3}_{-2.5}\times 10^{-3}$
	 &$1.1^{+0.2}_{-0.2}\times 10^{-3}$  &260 & 0.24 &  &  \\
HCG~62 $\dotfill$  & $2.0^{+0.2}_{-0.1}\times 10^{-3}$ & $6.4^{+0.2}_{-0.4}\times 10^{-3}$
	 &$1.0^{+0.2}_{-0.1}\times 10^{-3}$  &120 & 0.11 & $\sim 1.5$ &   \citet{tokoi08} \\
                   & $4.6^{+0.7}_{-0.6}\times 10^{-3}$ & $3.8^{+2.7}_{-3.4}\times 10^{-2}$
	 &$1.5^{+0.4}_{-0.4}\times 10^{-3}$  &230 & 0.21 & &   \\
A~262 $\dotfill$   & $3.6^{+0.1}_{-0.1}\times 10^{-3}$ & $1.2^{+0.3}_{-0.4}\times 10^{-2}$
	 &$1.6^{+0.2}_{-0.2}\times 10^{-3}$  &130 & 0.10 & $\sim 2$ &   \citet{sato09b} \\
                   & $6.7^{+0.4}_{-0.4}\times 10^{-3}$ & $3.7^{+1.2}_{-1.2}\times 10^{-2}$
	 &$2.7^{+0.7}_{-0.6}\times 10^{-3}$  &340 & 0.27 & &  \\
A~1060 $\dotfill$   & $5.7^{+0.4}_{-0.4}\times 10^{-3}$ & $4.3^{+0.8}_{-0.8}\times 10^{-2}$
	 &$2.4^{+0.5}_{-0.5}\times 10^{-3}$  &180 & 0.12 & $\sim 3$ &    \citet{sato07a} \\
                    & $4.0^{+0.4}_{-0.4}\times 10^{-3}$ & $4.3^{+2.0}_{-1.8}\times 10^{-2}$
	 &$1.6^{+0.8}_{-0.7}\times 10^{-3}$  &380 & 0.25 & &  \\
AWM~7 $\dotfill$   & $4.8^{+0.2}_{-0.2}\times 10^{-3}$ & $2.6^{+0.8}_{-0.8}\times 10^{-2}$
	 &$3.4^{+0.5}_{-0.5}\times 10^{-3}$  & 180 & 0.11 & $\sim 3.5$ &  \citet{sato08}\\
                   & $7.6^{+0.4}_{-0.3}\times 10^{-3}$ & $3.1^{+1.9}_{-1.2}\times 10^{-2}$
	 &$6.7^{+1.1}_{-1.1}\times 10^{-3}$  & 360 & 0.22 & &  \\
\hline
XMM-Newton & & & & & & & \\
Centaurus $\dotfill$ & $4\times 10^{-3}$ & $3\times 10^{-2}$ &-- &190 & 0.11 & $\sim 4$ &        \citet{matsushita07a} \\
\hline
\end{tabular}
\end{center}
\end{table*}

\begin{figure*}
\centerline{
\FigureFile(0.45\textwidth,1cm){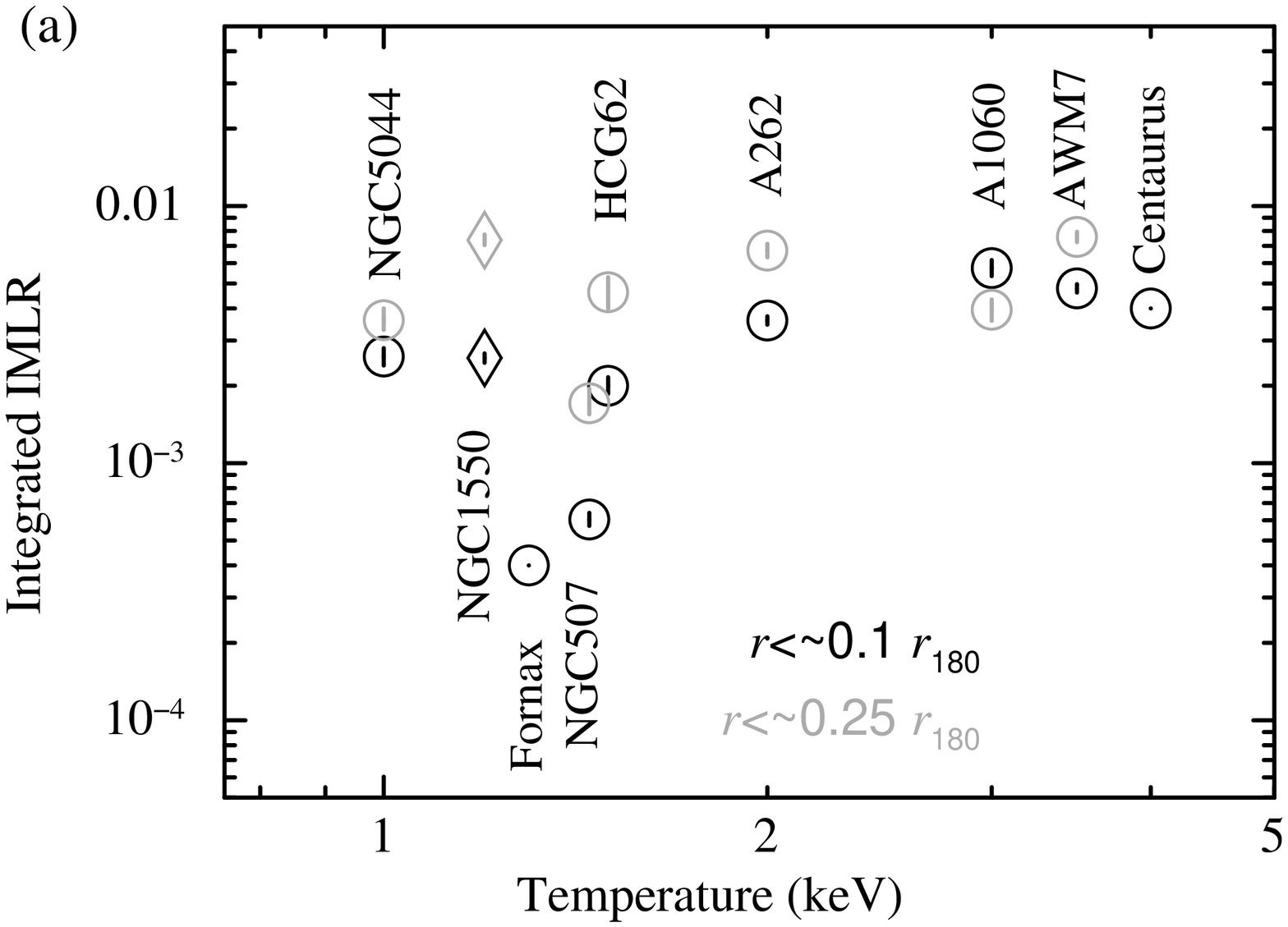}%
\hspace*{0.05\textwidth}
\FigureFile(0.45\textwidth,1cm){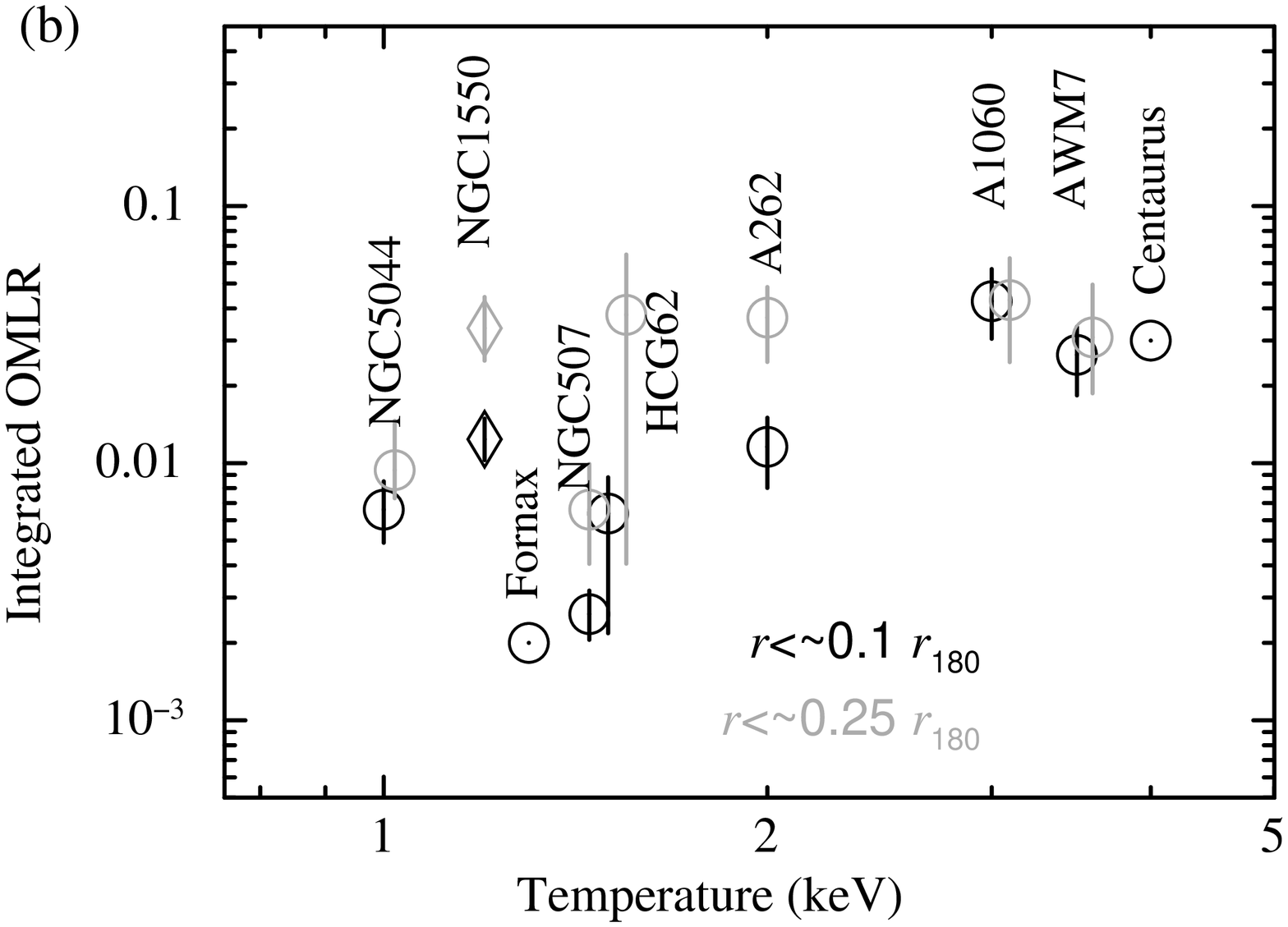}%
}
\caption{
Comparison of IMLR (a) and OMLR (b) 
with B-band luminosity to the other clusters and groups within 
$\sim0.1~r_{180}$ (black) and $\sim0.25~r_{180}$ (light gray) region. 
The MLRs of NGC~1550 are shown by the diamonds.
}\label{fig:6}
\end{figure*}

\subsection{Metal Mass-to-Light Ratio}

We derived 3-dimensional gas mass profile by extending the previous
XMM-Newton result \citep{kawaharada09} for the region within $14'$
arcmin ($\sim0.2~r_{180}$).  Combining it with the abundance
profiles obtained with Suzaku, we calculated cumulative metal mass.
The derived masses of Fe and Mg within the 3-dimensional
radius of $r< 457$~kpc ($r\sim0.5~r_{180}$) are $1.3\times 10^{9}$,
$3.4\times 10^{8}~M_\odot$, respectively, and the O mass within
$r<183$~kpc is $1.7\times 10^{9}$ $M_\odot$\@.  Errors of the metal
mass, which were used to calculate mass-to-light rations, were taken
from the statistical errors of each elemental abundance in the
spectral fits, because these are much larger than the error of gas
mass by \citet{kawaharada09}.

We examined mass-to-light ratios for O, Fe, and Mg
(OMLR, IMLR, and MMLR, respectively) which enabled us to compare the
ICM metal distribution with the stellar mass profile.  Historically,
B-band luminosity has been used for the estimation of the stellar mass
\citep{makishima01}, however we calculated it using the K-band
luminosity in NGC~1550 based on the Two Micron All Sky Survey (2MASS)
catalogue \footnote{ The database address: {\tt
    http://www.ipac.caltech.edu/2mass/}}. This method is useful in
performing a uniform comparison with the properties in other groups
and clusters based on the same K-band galaxy catalogue to trace the
distribution of member elliptical galaxies.

In the 2MASS catalog, we used all the data in a $2^{\circ}\times2^{\circ}$
region around NGC~1550 without the selection of galaxy morphology,
and subtracted the luminosity in a $r>1^{\circ}$ region, 
which corresponds to about $r_{\rm 180}$, 
as the background.  We then deprojected the
luminosity profile as a function of radius assuming a spherical
symmetry.  In order to convert the K-band magnitude of each galaxy to
the B-band value, we assumed the luminosity distance $D_{\rm
  L}=53.6$~Mpc, and an appropriate color $B-K=4.2$ for early-type
galaxies given by \citet{lin04}, along with the Galactic extinction
$A_B=0.583$ from NASA/IPAC Extragalactic Database (NED) in the
direction of NGC~1550\@.

The integrated values of OMLR, IMLR, and MMLR using the estimated
B-band luminosity within $r\lesssim 183$~kpc ($r\lesssim 0.2~r_{180}$)
turned out to be $\sim 3.3\times 10^{-2}$, $\sim 7.4\times 10^{-3}$,
and $\sim 3.3\times 10^{-3}$ $M_{\odot}/L_{\rm B \odot}$,
respectively, as shown in table~\ref{tab:5}.  The errors are based
only on the statistical errors of metal abundance in the spectral fit,
and the uncertainties in the gas mass profile and the luminosities of
member galaxies are not included. Note that we did not
adjust metal-mass and K-band profiles by considering the Suzaku PSF
effect, because uncertainties in the metal mass had the dominant
effect in our MLR estimation.

We compared these B-band MLRs for NGC~1550 with those of other groups
and clusters. The MLRs are all measured within inner
($\sim0.1~r_{180}$) and outer ($\sim0.25~r_{180}$) regions as shown in
table~\ref{tab:5} and figure~\ref{fig:6}. As for Fe (IMLR), NGC~1550
shows a similar value with the other groups and poor clusters in the
inner region.  As mentioned in subsection \ref{subsec:metal}, the Fe
abundance itself of NGC~1550 in the inner region is slightly smaller
than those in other groups and clusters.  However, the IMLR in this
region is almost the same as others, due to somewhat low value of
stellar mass in NGC~1550.  Looking at the outer region, NGC~1550 shows
slightly higher IMLR than NGC~5044 and comparable to those in other
poor clusters.  On the other hand for the OMLR in the inner region,
the poor systems show much lower values than the larger
high-temperature systems. Interestingly, the poor systems (including
NGC~1550) show higher OMLR in the outer region, comparable to those in
the larger systems. The spatial extent of O looks to be relatively
large in these very small systems.

\citet{rasmussen09} suggests that the IMLRs within $r_{500}$ of 
15 groups of galaxies observed with Chandra have a positive 
correlation with the groups mass (temperature). Our results 
and Chandra's results are also almost consistent, and the 
IMLRs of groups are slightly lower than those of clusters. 

\begin{table*}
\caption{
Integrated mass-to-light ratios of O, Mg, and Fe (OMLR, MMLR, IMLR ) 
with K-band luminosity in units of $M_{\odot}/L^{\rm K}_{\odot}$.
}\label{tab:6}
\begin{center}
\begin{tabular}{lccc}
\hline
\multicolumn{1}{c}{Region (kpc/$r_{180}$)} & OMLR & MMLR & IMLR \\
\hline
$<$45.7/0.05 & 9.4$^{+2.2}_{-2.1}\times10^{-4}$ &
 1.1$^{+0.1}_{-0.1}\times10^{-4}$ & 2.9$^{+0.1}_{-0.1}\times10^{-4}$ \\
$<$91.3/0.09 & 3.5$^{+0.7}_{-0.6}\times10^{-3}$ &
 2.6$^{+0.3}_{-0.3}\times10^{-4}$ & 7.1$^{+0.3}_{-0.3}\times10^{-4}$ \\
$<$182.6/0.19& 9.2$^{+0.3}_{-0.3}\times10^{-3}$ &
 9.0$^{+1.4}_{-1.3}\times10^{-4}$ & 2.0$^{+0.1}_{-0.1}\times10^{-3}$ \\
$<$456.6/0.47& -- &
 9.0$^{+1.0}_{-4.5}\times10^{-4}$ & 3.5$^{+0.5}_{-0.5}\times10^{-3}$ \\
\hline
\end{tabular}
\end{center}
\end{table*}

\begin{figure*}
\centerline{
\FigureFile(0.45\textwidth,1cm){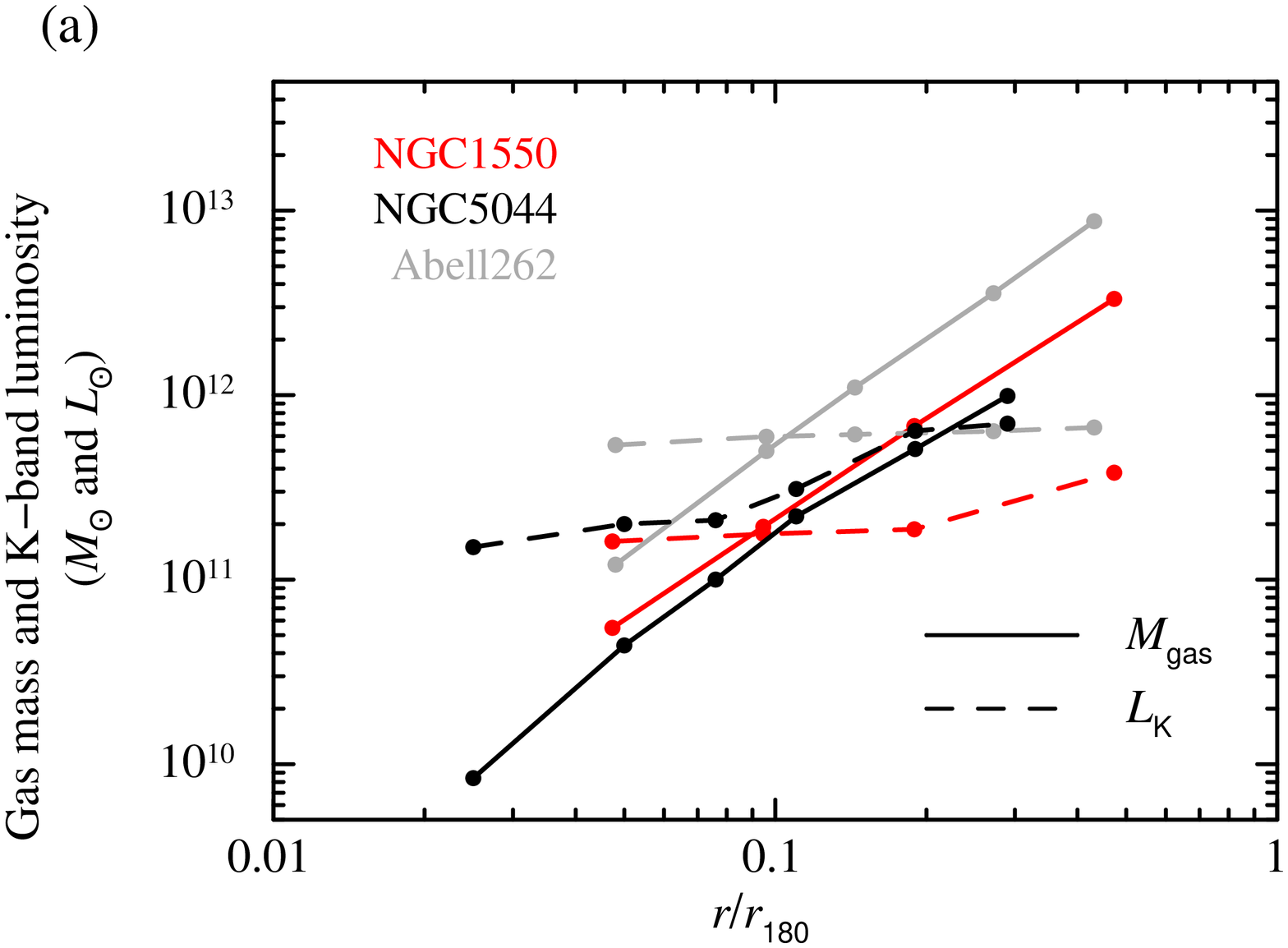}%
\hspace*{0.05\textwidth}
\FigureFile(0.45\textwidth,1cm){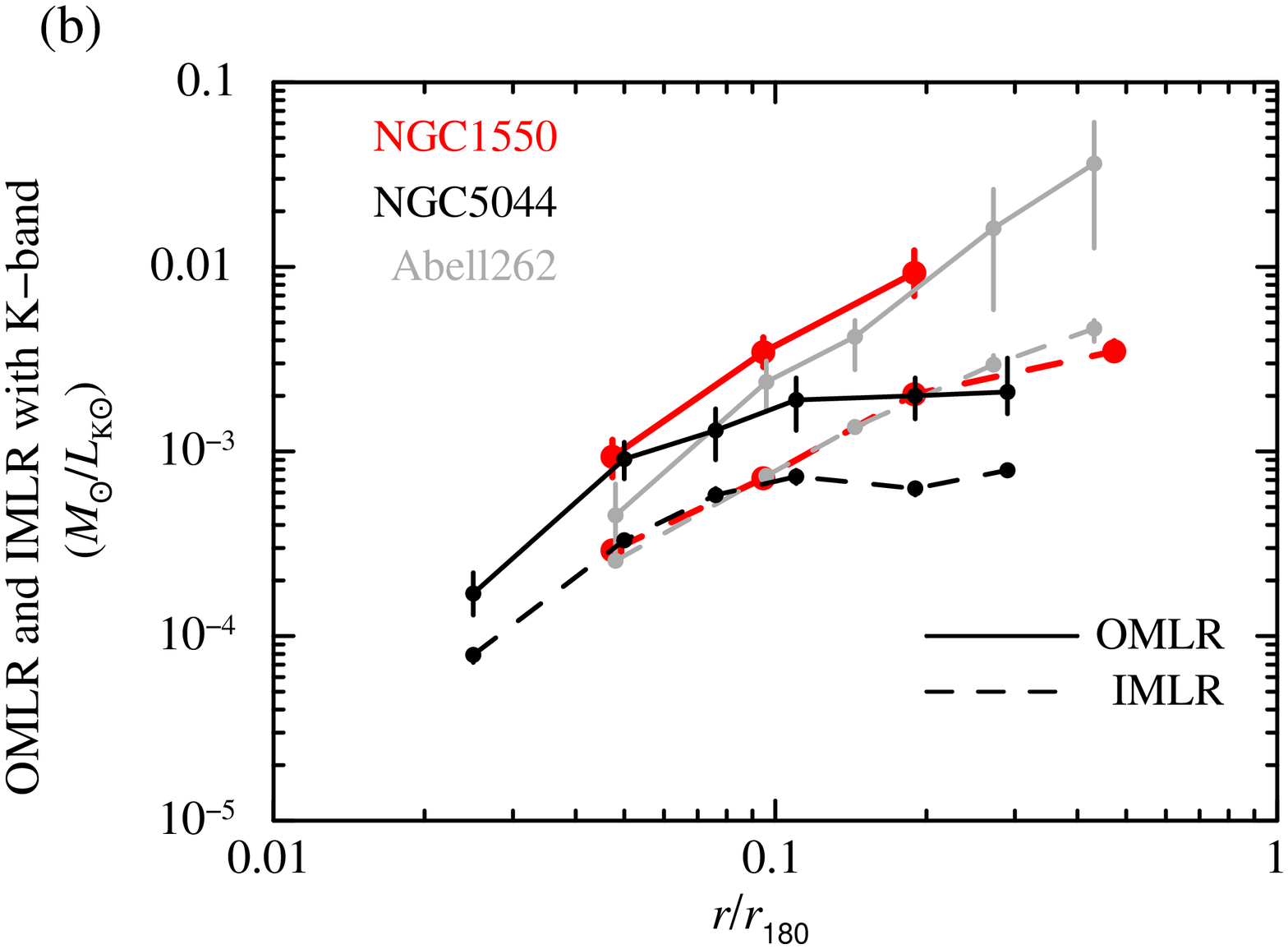}%
}
\caption{
(a) Gas mass, $M(<R)$, and K-band luminosity, $L_{\rm K}$ 
within the 3-dimensional radius, $R$, in NGC~1550, NGC~5044, and Abell~262.
The gas mass profiles derived from XMM-Newton observations 
for NGC~1550 \citep{kawaharada09}, NGC~5044 \citep{komiyama09}, 
and Abell~262 \citep{gastaldello07}. 
The red, black, and light-gray lines show each value for NGC~1550, NGC~5044, 
and Abell~262, respectively. The solid and dashed lines corresponds 
to each gas mass and K-band luminosity, respectively.
(b) Ratios of the O and Fe mass in units of $M_\odot$
to the $K$ band luminosity in units of $L_\odot$
(OMLR and IMLR, respectively) for NGC~1550, NGC~5044, and 
Abell~262, against the 3-dimensional radius. 
The solid and dashed lines corresponds to each OMLR and IMLR, respectively.
}\label{fig:7}
\end{figure*}

We also calculated the MLRs by directly using the K-band luminosity
assuming the Galactic extinction $A_K=0.050$ in the direction of
NGC~1550, and the absolute K-band solar magnitude of 3.34\@.  The
resultant K-band luminosity within the Suzaku observed region,
$r<30'$, is $3.8\times10^{11}~L_{\rm K\odot}$, and the radial
luminosity profile is also plotted in figure ~\ref{fig:7}(a).  We
calculated the radial profile of the OMLR, IMLR, and MMLR values using
the K-band luminosity out to a radius $r \sim 180$~kpc ($r\lesssim
0.2~r_{180}$), as shown in figure~\ref{fig:7}(b) and
table~\ref{tab:6}. The values at the outermost radius are $\sim
9.2\times 10^{-3}$, $\sim 2.0\times 10^{-3}$, and $\sim 9.0\times
10^{-4}$ $M_{\odot}/L_{\rm K\odot}$, respectively.

In order to investigate the dependence on the system size, we compared
the MLRs of NGC~1550 ($kT \approx 1.2$ keV) with those of NGC~5044 (1
keV) and Abell~262 (2 keV), which are all poor relaxed cooling-core
groups and clusters.  Although they show similar IMLR in the inner
region, NGC~1550 and Abell~262 show gas mass ratio to the K-band
luminosity higher than NGC~5044, as shown in figure~\ref{fig:7}(a).
As one goes outside ($r>0.1~r_{180}$), NGC~1550 shows fairly
consistent MLRs with Abell~262, larger than those in NGC~5044 as shown
in figure~\ref{fig:6}(b).  The radial profiles of IMLR and OMLR for
NGC~1550 look quite similar to those for Abell~262 rather than
for NGC~5044.

The radius, $r\sim0.1~r_{180}$, corresponds to the region where the
gas mass and K-band luminosity seem to overlap as shown in
figure~\ref{fig:7}(a).  
Assuming the stellar mass-to-light ratio,
$M_{\rm star}/L_{\rm K} \sim 1$ as shown in \citet{arnouts07}
(see also \cite{nagino09}), this
radius indeed corresponds to the point where the gas and stellar
masses are comparable.  The similarity of MLRs for the 3 systems
examined here in the inner region ($r<0.1~r_{180}$) suggests that the
metal enrichment within the past few Gyr has occurred in a similar
way.  On the other hand, in the outer region ($r>0.1~r_{180}$),
NGC~5044 shows lower IMLR than those in NGC~1550 and Abell~262,
even though Fe abundances themselves are comparable in these regions.
This suggests that distribution of stellar mass has some different
history between these systems.

We will consider how the observed features of abundance and MLR
profiles can constrain enrichment scenario. First, metals in the inner
region ($r \lesssim 0.1~r_{180}$) have been mostly supplied by the
central galaxy, so this region should be set aside in the present
discussion.  In the immediate outer region ($r\gtrsim 0.1~r_{180}$),
metals from the central galaxy could not reach in a few Gyr time
scale. Also, we may assume that almost the same amount of metals per
stellar mass was synthesized in all systems before the collapse of
groups and clusters. In this case, we expect very similar MLRs in
different systems, contrary to the observed feature.  As shown in
figure \ref{fig:7}(b), at least NGC~5044 shows lower IMLR and OMLR
profiles compared with those for two other systems.  This implies that
the thermal and/or dynamical evolutions of the gas have different
history among different systems during or after the collapsing period
of individual groups and clusters.

\citet{renzini05} showed the expected MLRs as a function of the IMF slope.
As for OMLR, the expected value to be $\sim0.1~M_{\odot}/L_{\rm B}$
at a Salpeter IMF is slightly higher than our results
of $\sim0.03~M_{\odot}/L_{\rm B}$ within $\sim0.25~r_{180}$ 
as shown in figure \ref{fig:6}(b). 
However, as shown in figure \ref{fig:7}(b), because the OMLRs 
look increasing toward outer region, the OMLRs within the virial radius
would be represented with the Salpeter IMF. \citet{renzini05} also 
suggested that a top heavy IMF would overproduce metals by more than 
a factor of 20, which is much larger than the observed values 
including our results.

We stress that high-sensitivity abundance observation to the outer
region of clusters will give important clues about their evolution.
If O distribution, as well as Fe, could be measured to the very outer
region ($r\sim r_{180}$), we may obtain a clear view about when O and
Fe were supplied to the inter galactic space because most of O should
have been synthesized by SNe II and supplied in the starburst era.
Another possibility is very early metal enrichment of O by galaxies or
massive Population III stars (e.g.\cite{matteucci06}) before groups
and clusters assemble. In this case, a large part of the intergalactic
space would be enriched quite uniformly with O and other elements.
Metallicity information in cluster outskirts would thus give us unique
information about the enrichment history.  For this purpose,
instruments with much higher energy resolution, such as
microcalorimeters, and optics with larger effective area will play a
key role in carrying out these studies.

\section{Summary and conclusion}

Suzaku observation of the fossil group NGC~1550 showed spatial
distributions of temperature and metal abundances for O, Mg, Si, S,
and Fe up to $\sim0.5~r_{180}$ for the first time, and 
confirmed that the metals in the ICM of this fossil group are indeed 
extending to a large radius. The ICM temperature
decreases mildly from $\sim1.5$ keV to $\sim1.0$ keV in the outer
region, similar to the feature seen in other clusters.  The abundances
of Mg, Si, S, and Fe drop from subsolar levels at the center to $\sim
1/4$ solar in the outermost region, while the O abundance shows a
flatter distribution around $\sim 0.5$ solar without the strong
concentration in the center.  The abundance ratios, O/Fe, Mg/Fe,
Si/Fe, and S/Fe for NGC~1550 are generally similar to those in
groups and poor clusters.  The abundance pattern from O to Fe enabled us to
constrain number ratio of SNe II to Ia as $2.9 \pm 0.5$, which is
consistent with the values obtained for other groups and clusters.
The derived MLRs of NGC~1550 using the B-band and K-band luminosities
are consistent with those in the NGC~5044 group for the inner region
$r\lesssim 0.1~r_{180}$, while NGC~1550 shows slightly higher MLRs than
NGC~5044 in the outer region. This suggests that metal enrichment
process may reflect the size of the system in the sense that larger
systems contain higher amount of metals for a given stellar mass.

\bigskip
Authors thank the referee for providing valuable comments.
K.S is supported by a JSPS Postdoctral fellowship for research abroad.
Part of this work was financially supported
by the Ministry of Education, Culture, Sports, Science
and Technology, Grant-in-Aid for Scientific Research Nos.
20340068, 21224003, 21740134.

\end{document}